\documentclass[numberedappendix]{emulateapj}

\usepackage{epsf}
\usepackage{lscape}

\newcommand{\noi}{\smallskip\noindent}

\slugcomment{Accepted by ApJ}

\shorttitle{}
\shortauthors{}

\begin{document}

\title{The Sloan Lens ACS Survey. III -- the structure and formation
of\\ early-type galaxies and their evolution since $z\approx1$}

\author{ L\'eon V.E.\ Koopmans\altaffilmark{1},
  Tommaso\ Treu\altaffilmark{2},
  Adam S.\ Bolton\altaffilmark{3,5},
  Scott\ Burles\altaffilmark{3}
  and Leonidas A.\ Moustakas\altaffilmark{4}}

\altaffiltext{1}{Kapteyn Astronomical Institute, University of
Groningen, P.O.Box 800, 9700 AV Groningen, The Netherlands ({\tt
koopmans@astro.rug.nl})}

\altaffiltext{2}{Department of Physics, University of California,
Santa Barbara, CA 93106, USA ({\tt ttreu@physics.ucsb.edu})}

\altaffiltext{3}{Department of Physics and Kavli Institute for
Astrophysics and Space Research, Massachusetts Institute of
Technology, 77 Massachusetts Ave., Cambridge, MA 02139, USA ({\tt
burles@mit.edu})}

\altaffiltext{4}{Jet Propulsion Laboratory, Caltech, MS 169-327, 4800
Oak Grove Dr., Pasadena, CA 91109 ({\tt leonidas@jpl.nasa.gov})}

\altaffiltext{5}{Harvard-Smithsonian Center for Astrophysics,
60 Garden St. MS-20, Cambridge, MA 02138 USA ({\tt abolton@cfa.harvard.edu})}

\begin{abstract}

We present a joint gravitational lensing and stellar dynamical
analysis of fifteen massive field early-type galaxies -- selected from
the {\sl Sloan Lens ACS} (SLACS) Survey -- using {\sl Hubble Space
Telescope} ACS images and luminosity weighted stellar velocity
dispersions obtained from the Sloan Digital Sky Survey database.  The
sample of lens galaxies is well-defined (see Paper I), with a redshift
range of $z$=0.06--0.33 and an average stellar velocity dispersion of
$\langle \sigma_{\rm ap} \rangle = 263$\,km\,s$^{-1}$ (rms of
44\,km\,s$^{-1}$) inside a 3-arcsec fiber diameter.
The following numerical results are found: (i) A joint-likelihood
gives an average logarithmic density slope for the {\sl total} mass
density of $\langle \gamma' \rangle = 2.01^{+0.02}_{-0.03}$ (68\%
C.L.; $\rho_{\rm tot}\propto r^{-\gamma'}$) inside $\langle {\rm
R}_{\rm Einst} \rangle = 4.2 \pm 0.4$\,kpc (rms of 1.6\,kpc). The
inferred {\it intrinsic} rms spread in logarithmic density slopes is
$\sigma_{\gamma'}=0.12$, which might still include some minor
systematic uncertainties. A range for the stellar anisotropy parameter
$\beta=[-0.25, +0.25]$ results in $\Delta\langle
\gamma'\rangle=[+0.05,-0.09]$. Changing from a Hernquist to a Jaffe
luminosity density profile increases $\langle \gamma' \rangle$
by~0.05. (ii) The average position-angle difference between the light
distribution and the total mass distribution is found to be $\langle
\Delta \theta\rangle = 0 \pm 3$~degrees (rms of 10 degrees), setting
an upper limit of $\langle \gamma_{\rm ext}\rangle \la 0.035$ on the
average external shear.  The total mass has an average ellipticity
$\langle q_{\rm SIE}\rangle$=0.78$\pm$0.03 (rms of 0.12), which
correlates extremely well with the stellar ellipticity, $q_*$,
resulting in $\langle q_{\rm SIE}/q_* \rangle = 0.99 \pm 0.03$ (rms of
0.11) for $\sigma \ga 225$~km\,s$^{-1}$. At lower velocity
dispersions, inclined S0 galaxies dominate, leading to a higher ratio
(up to 1.6). This suggests that the dark-matter halo surrounding these
galaxies is less flattened than their stellar component. Assuming an
oblate mass distribution and random orientations, the distribution of
ellipticities implies $\langle q_3 \rangle\equiv
\langle(c/a)_\rho\rangle =0.66$ with an error of $\sim$0.2.  (iii) The
average projected dark-matter mass fraction is inferred to be $\langle
f_{\rm DM} \rangle =0.25 \pm 0.06$ (rms of 0.22) inside $\langle {\rm
R}_{\rm E}\rangle$, using the stellar mass-to-light ratios derived
from the Fundamental Plane as priors.  (iv) Combined with results from
the {\sl Lenses Structure \& Dynamics} (LSD) Survey at $z \ga 0.3$, we
find no significant evolution of the total density slope inside one
effective radius for galaxies with $\sigma_{\rm ap}\ge
200$~km\,s$^{-1}$: a linear fit gives $\alpha_{\gamma'} \equiv
d\langle \gamma' \rangle/dz=0.23\pm0.16$ (1\,$\sigma$) for the range
$z$=0.08--1.01.
We conclude that massive early-type galaxies at $z$=0.06--0.33 on
average have an isothermal logarithmic density slope inside half an
effective radius, with an intrinsic spread of at most 6\,\%
(1\,$\sigma$). The small scatter and absence of significant evolution
in the inner density slopes suggest a collisional scenario where gas
and dark matter strongly couple during galaxy formation, leading to a
total mass distribution that rapidly converge to dynamical {\sl
isothermality}.

\end{abstract}
\keywords{gravitational lensing --- galaxies: elliptical and
lenticular, cD --- galaxies: evolution --- galaxies: formation ---
galaxies: structure}

\section{Introduction}

Massive early-type galaxies are postulated to be latecomers in the
hierarchical formation process (e.g. Blumenthal et al.\ 1984; Frenk et
al.\ 1985), formed via mergers of lower-mass (disk) galaxies (e.g.\
Toomre \& Toomre 1972; Schweizer 1982; Frenk et al.\ 1988; White \&
Frenk 1991; Barnes 1992; Cole et al. 2000). As such, a detailed study
of their structure (e.g.\ Navarro, Frenk \& White 1996; Moore et al.\
1998), formation and subsequent evolution provides a powerful test of
the concordance $\Lambda$CDM paradigm (e.g.\ Riess et al.\ 1998;
Perlmutter et al.\ 1999; Spergel et al.\ 2003; Tegmark et al.\ 2004)
at galactic scales.

In this context, the merging of low-mass galaxies to form more massive
ones naively seems to imply a continuous evolution of their mass
structure (e.g.\ Bullock et al.\ 2001), both in their outer regions
and dense inner regions (e.g.\ smaller galaxies accrete and sink to
the center through dynamical friction). On the one hand, the inner
regions of massive ellipticals can contract into an increasingly
denser structure, if significant mass in dissipational gas is accreted
(e.g.\ Blumenthal et al.\ 1986; Ryden \& Gunn 1987; Navarro \& Benz
1991; Dubinski 1994; Jesseit, Naab \& Burkert 2002; Gnedin et al.\
2004; Kazantzidis et al.\ 2004). If this process occurs at
$z$\,$\la$\,1 (e.g.\ Kauffmann, Charlot \& White 1996; Kauffmann \&
Charlot 1998) and results in star-formation activity, one can test
this scenario directly by using high-quality data of early-type
galaxies, obtained with space and 8--10\,m class ground-based
telescopes (e.g.\ Menanteau et al.\ 2001a\&b; Gebhardt et al. 2003;
McIntosh et al.\ 2005; Tran et al.\ 2005). On the other hand, the mass
inside the inner $\sim$10\,kpc of the most massive ellipticals (i.e.\
$>$L$_*$) seems to remain nearly constant from $z\gg1$ to the present
day -- as suggested by collisionless dark-matter simulations -- and
additionally accreted dark matter replaces already-present
collisionless matter (e.g.\ Wechsler et al.\ 2002; Zhao et al.\
2003; Gao et al.\ 2004). If most gas is turned into (collisionless)
stellar mass before mergers (e.g. van Dokkum et al. 1999), one expects
it to behave similarly to dark matter during assembly in to the more
massive galaxies seen at $z \la 1$.

Based on the notion that the velocity function of massive early-type
galaxies at $z$=0 (Sheth et al.\ 2003) is remarkably close to that of
the inner regions (inside $\sim$10~kpc) of the most massive simulated
galaxies at $z$\,$\approx$\,6 -- even though these galaxies continue
to accrete collisionless matter below that redshift -- Loeb \&
Peebles\ (2003) suggest that the inner regions might behave as {\sl
dynamical attractors}, whose phase-space density is nearly invariant
under the accretion of collisionless matter (see also e.g.\ Gao et
al.\ 2004; Kazantzidis, Zentner \& Kravtsov 2005). In this scenario,
one might expect less structural evolution of the inner regions of
massive early-type galaxies at $z<1$, compared to models where most
gas had not yet turned into stars {\sl before} the mass assembly of
their inner regions took place.  Hence, one way to study the formation
scenario of massive ellipticals, is to quantify the evolution of the
mass distribution in their inner regions from redshifts $z$=1 to~0.

\begin{figure*}[t]
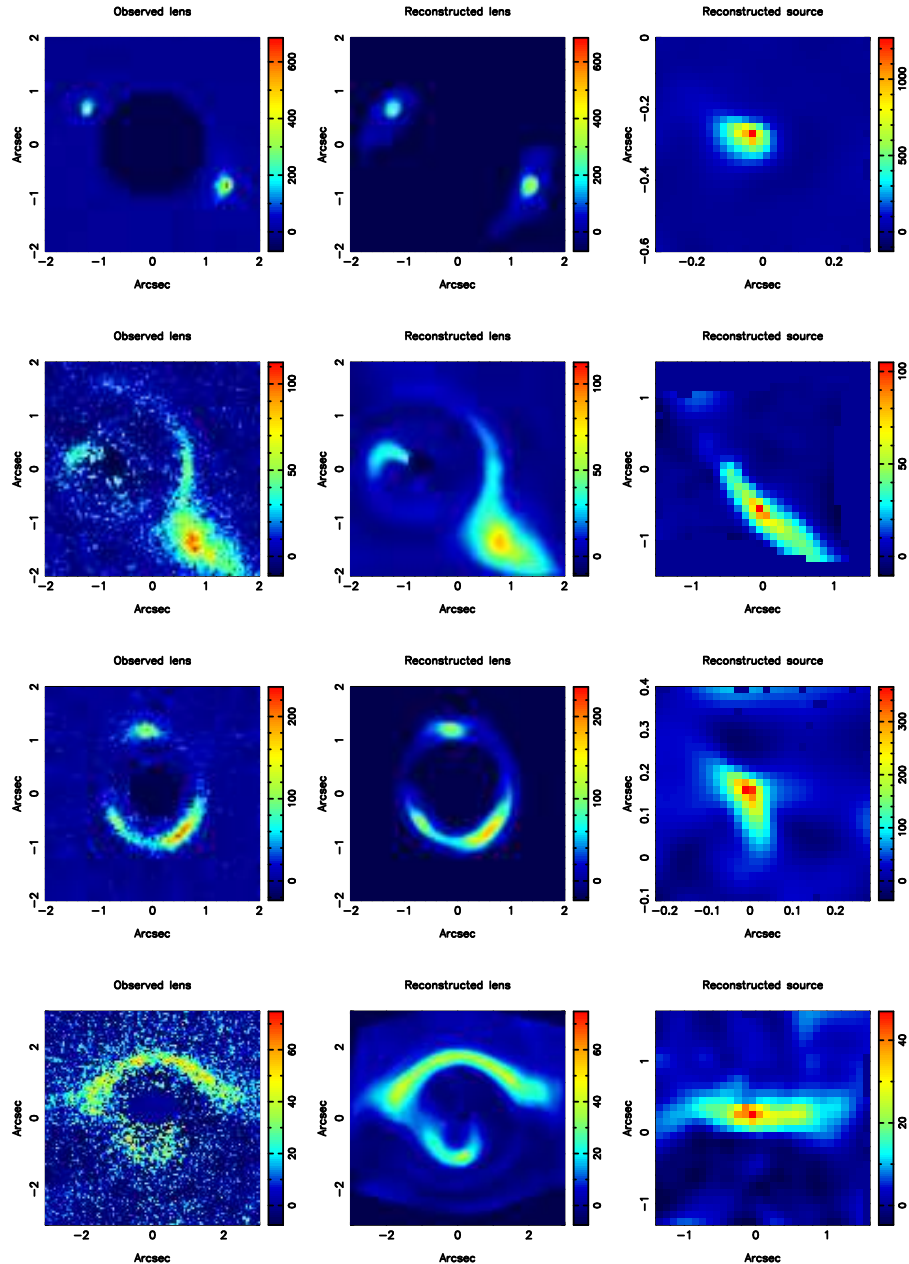

\begin{center}
\leavevmode
\vbox{%
\vbox{%
\epsfxsize=0.70\hsize
\epsffile{fig1a.epsi}
\epsfxsize=0.70\hsize
\epsffile{fig1b.epsi}
\epsfxsize=0.70\hsize
\epsffile{fig1c.epsi}
\epsfxsize=0.70\hsize
\epsffile{fig1d.epsi}}
}
\end{center}
\caption{Non-parametric lens-image reconstructions of confirmed SLACS
lens systems. For each system, the observed (galaxy-subtracted)
HST-ACS F814W image is shown (left panel), the best reconstruction of
the system (middle panel) and source model (right panel), assuming a
SIE mass model. From top to bottom are shown: J0037$-$0942,
J0216$-$0813, J0737+3216, and J0912+0029 (see Paper I and
Table~\ref{tab:results}).}
\label{fig:siemodels}
\end{figure*}

\addtocounter{figure}{-1}

\begin{figure*}[t]
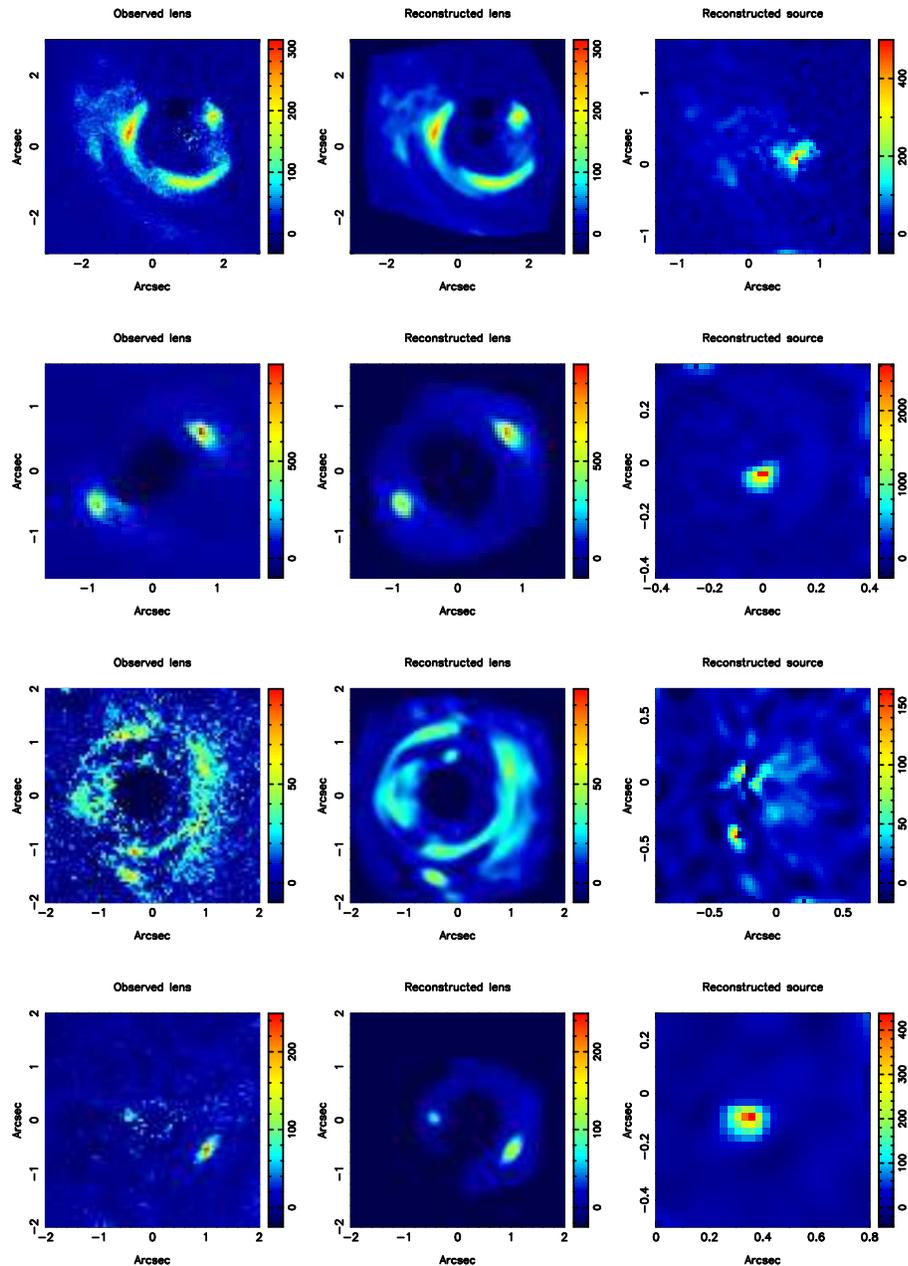

\begin{center}
\leavevmode
\vbox{%
\epsfxsize=0.70\hsize
\epsffile{fig1e.epsi}
\epsfxsize=0.70\hsize
\epsffile{fig1f.epsi}
\epsfxsize=0.70\hsize
\epsffile{fig1g.epsi}
\epsfxsize=0.70\hsize
\epsffile{fig1h.epsi}
}
\end{center}
\caption{(Continued) From top to bottom are shown: J0956+5100,
J0959+0410, J1250+0523 and J1330$-$0148.}
\end{figure*}

\addtocounter{figure}{-1}

\begin{figure*}[t]
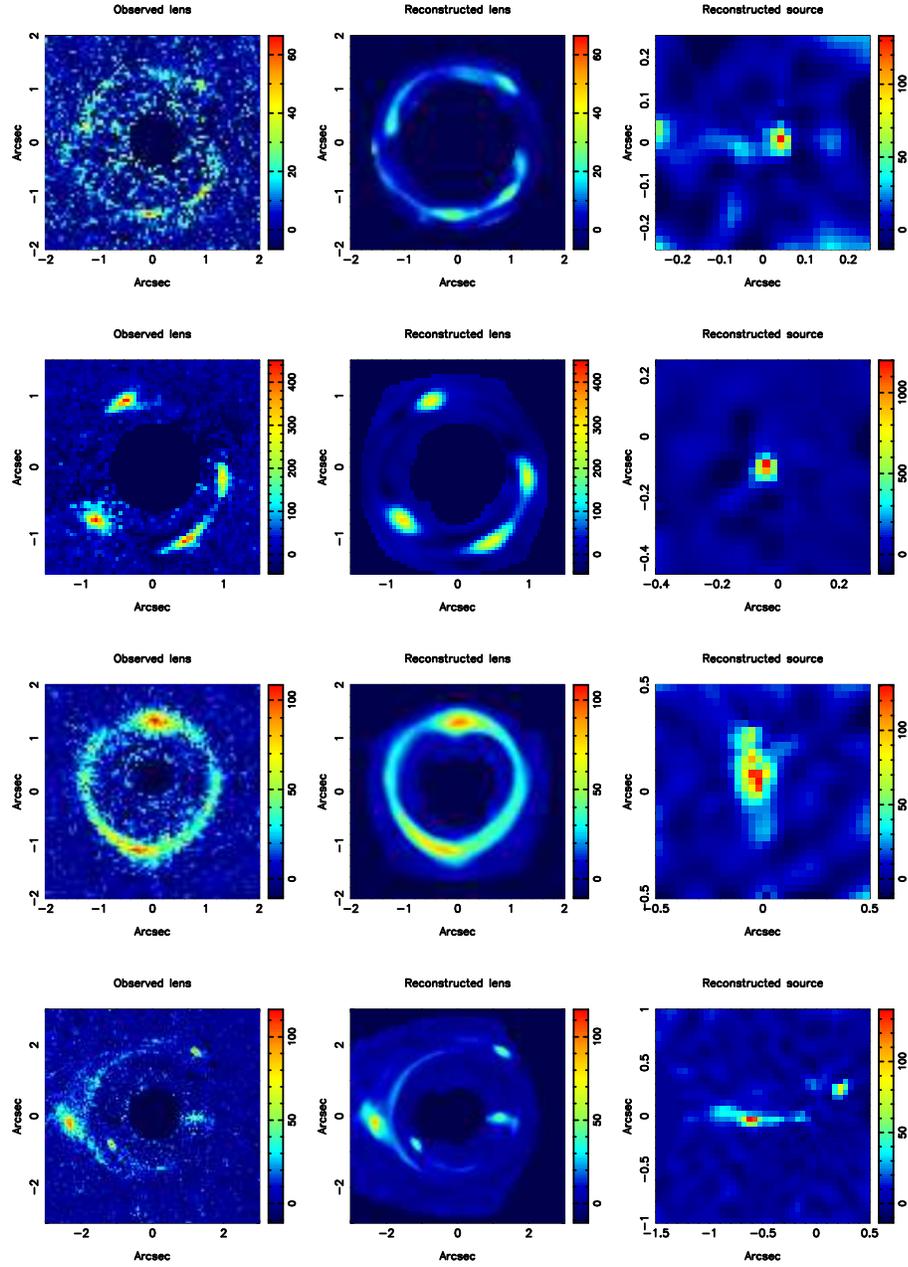

\begin{center}
\leavevmode
\vbox{%
\epsfxsize=0.70\hsize
\epsffile{fig1i.epsi}
\epsfxsize=0.70\hsize
\epsffile{fig1j.epsi}
\epsfxsize=0.70\hsize
\epsffile{fig1k.epsi}
\epsfxsize=0.70\hsize
\epsffile{fig1l.epsi}
}
\end{center}
\caption{(Continued) From top to bottom are shown: J1402+6321,
J1420+6019, J1627$-$0055 and J1630+4520.}
\end{figure*}

\addtocounter{figure}{-1}

\begin{figure*}[t]
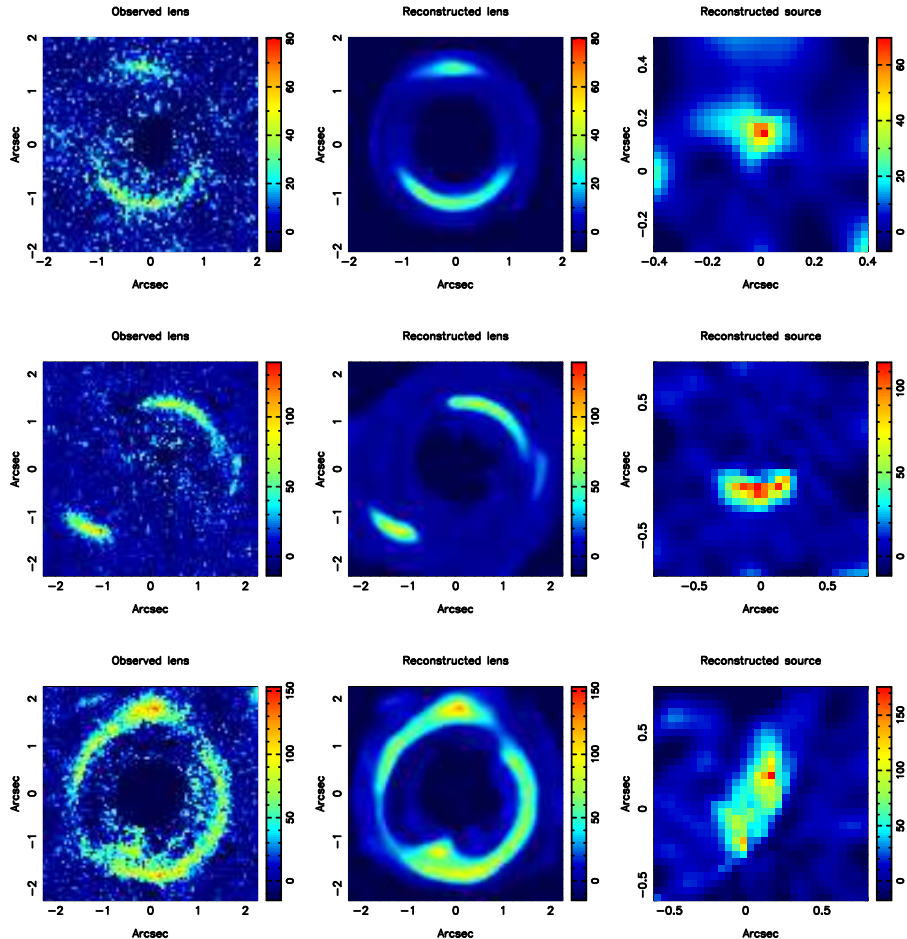

\begin{center}
\leavevmode
\vbox{%
\epsfxsize=0.70\hsize
\epsffile{fig1m.epsi}
\epsfxsize=0.70\hsize
\epsffile{fig1n.epsi}
\epsfxsize=0.70\hsize
\epsffile{fig1o.epsi}
}
\end{center}
\caption{(Continued) From top to bottom are shown: J2300+0022,
J2303+1422 and J2321$-$0939.}
\end{figure*}

From the observational point of view, a significant effort has been
devoted in the past two decades to the study of the mass structure of
early-type galaxies in the local Universe ($z$$\la$0.1) through
stellar dynamical tracers and X-ray studies (e.g.\ Fabbiano 1989;
Mould et al. 1990; Matsushita et al. 1998; Loewenstein \& White 1999;
Saglia et al. 1992; Bertin et al. 1994; Arnaboldi et al. 1996; Franx
et al. 1994; Carollo et al. 1995; Rix et al. 1997; Gerhard et
al. 2001; Seljak 2002; Borriello et al. 2003; Romanowsky et
al. 2003). In a comprehensive study, Gerhard et al.\ (2001) conclude
that massive ellipticals have, on average, flat circular velocity
curves with a scatter of $\sim$10\% in their inner two effective
radii. This in itself should impose stringent constraints on any
numerical simulations of elliptical galaxies (e.g.\ Meza et al.\ 2003;
Kawata \& Gibson 2003).

Strong gravitational lensing provides a complementary approach
(Kochanek 1991) to study early-type galaxies at higher
redshifts. Lensing analysis has been used to demonstrate the presence
of dark matter around early-type galaxies and, in some systems, to
provide evidence for ``isothermal'' (i.e. $\rho_{\rm tot}\propto
r^{-2}$) mass density profiles equivalent to the flat rotation curves
observed for spiral galaxies (e.g., Kochanek 1995; Rusin \& Ma 2001;
Ma 2003; Rusin et al. 2002, 2003a; Cohn et al. 2001; Munoz et
al. 2001; Winn et al. 2003; Wucknitz et al. 2004; Rusin \& Kochanek
2005). However, the mass-profile (e.g.\ Wucknitz 2002) and mass-sheet
degeneracies (Falco et al.\ 1985) often prevent a truly accurate
determination of the logarithmic density slope {\sl at} the Einstein
radius.

To answer the question ``{\sl What is the mass structure inside the
inner regions of early-type galaxies and how does it evolve with
time?\,}'', we therefore combine constraints from strong gravitational
lensing and stellar kinematics. The former provides an accurate mass
measurement inside the Einstein radius (Kochanek 1991), whereas the
latter provides a measurement of the mass gradient. The average
logarithmic density slope inside the Einstein radius can then be
determined -- independent of the mass-sheet degeneracy that is
associated with the galaxy mass distribution -- with the same
fractional accuracy as is obtained on the luminosity-weighted stellar
velocity dispersion (e.g.\ see Treu \& Koopmans 2002 and Koopmans
2004).  The density slopes of {\sl individual} early-type galaxy can
be correlated with redshift, to determine any structural evolution in
the population.

In an ongoing study of massive early-type lens galaxies between $z
\approx 0.5$ and 1, as part of the {\sl Lenses Structure \& Dynamics}
(LSD) Survey (Koopmans \& Treu 2002, 2003; Treu \& Koopmans 2002,
2003, 2004; hereafter TK04) -- plus two additional systems that were
studied to measure the Hubble Constant (Treu \& Koopmans 2002;
Koopmans et al 2003) -- this technique has successfully been applied,
to place the first constraints on the inner density slopes and
dark-matter halos of early-type galaxies to $z\approx 1$ (TK04),
finding a logarithmic density slope close to isothermal, although the
results were limited by the small sample size.

In the first paper of this series (Bolton et al.\ 2006; hereafter
Paper I), we reported on the discovery of nineteen new early-type lens
galaxies from the {\sl Sloan Lens ACS} (SLACS) Survey at $z \la 0.3$,
each with {\sl Hubble Space Telescope} (HST) F435W and F814W images
and a stellar velocity dispersion measured from their SDSS spectra
(e.g.\ Bolton et al.\ 2004). Some systems have integral field
spectroscopy (IFS) of their lensed sources, obtained with Magellan
and/or Gemini (see also Bolton et al.\ 2005). As far as their
photometric properties are concerned, they are representative of
Luminous Red Galaxies (LRG; Eisenstein et al.\ 2004) with similar
redshifts and similar stellar velocity dispersions (see Paper I). They
also lie on the Fundamental Plane (FP; Dressler et al. 1987;
Djorgovski \& Davis 1987) of early-type galaxies, have old stellar
populations, and have very homogeneous mass density profiles (Treu et
al.\ 2006; hereafter Paper II).

In this paper, we focus on the analysis of a sub-sample of fifteen
isolated early-type lens galaxies from SLACS, combining the
constraints from {\sl Hubble Space Telescope} (HST) images with the
stellar velocity dispersion obtained from the SDSS database. The goals
are to quantify their inner mass structure and to assess whether any
evolution of their inner regions has occurred at~$ z \la 1$. In \S\,2,
we present non-parametric lens models for each system. The simplicity
of the models supports their lensed nature and provides the necessary
input for subsequent analysis. In \S\,3, we use the enclosed mass from
lensing in a joint stellar-dynamical analysis, to determine the inner
density slopes of each early-type galaxy.  In combination with results
from the LSD survey, we analyze the redshift behavior of the density
slope in \S\,4 to quantify its evolution. In \S\,5, we summarize our
results and draw conclusions. Throughout this paper, we assume
H$_0$=70\,km\,s$^{-1}$\,Mpc$^{-1}$, $\Omega_{\rm m}=0.3$ and
$\Omega_\Lambda = 0.7$.

\section{Gravitational-Lens Models}

In this section, we briefly summarize the selection procedure of lens
candidates for the HST snaphot program (see Bolton et al.\ 2004 and
Paper I) and the sub-sample of the fifteen early-type lens galaxies
that we use throughout this paper, in addition to their lens models.
The models are used to quantify the alignment between stellar and
total mass (\S\,2.4) and provide the mass enclosed by the lensed
images as an external constraint on the stellar-dynamical models
(\S\,3).

\subsection{The Sample}

The selection procedure that led to the current sample of confirmed
E/S0 lens galaxies, used in this paper, is as follows: first, a
principle-component analysis (PCA) is done of all spectra in the LRG
and MAIN galaxy samples from the SDSS. The smooth PCA spectra are
subtracted from the observed spectra and the residuals are studied for
absorption and higher-redshift emission lines. The absorption lines
secure the redshift of the foreground galaxy and allow the
luminosity-weighted stellar velocity dispersions to be measured for
the brighter galaxies. Second, the residual spectra that show three or
more atomic transition lines in emission, including [O\,{\footnotesize
II}], at a single redshift beyond that of the main galaxy are selected
for follow-up. Third, given the redshifts and stellar velocity
dispersions ($\sigma_{\rm ap}$) of the foreground galaxies, and the
redshifts of the lensed source candidates, we can estimate their
lensing probability by ranking the systems according to their Einstein
radii $\theta_{\rm E}=4\pi (\sigma_{\rm ap}/c)^2 D_{\rm ds}/D_{\rm
s}$, assuming a SIS mass model with $\sigma_{\rm SIS} \equiv
\sigma_{\rm ap}$ (e.g.\ Schneider et al.\ 1992) from large to
small\footnote{Note that in the context of the SIS model, all systems
with $2\,\theta_{\rm E} \ge 1.5''$ (the SDSS fiber radius) have a
probability of unity to be multiply imaged lens systems, because one
of the two lensed images forms between 0--1\,$\theta_{\rm E}$ and the
second image forms between 1--2\,$\theta_{\rm E}$. Hence, observing an
image inside 2\,$\theta_{\rm E}$ implies that a counter image must
exist.}.  This procedure resulted in a ranked list of 49 candidates,
of which 20 are from the LRG sample in Bolton et al.\ (2004) and the
remaining 29 are from the MAIN galaxy sample. We further note that
galaxies in the LRG sample were selected based on early-type spectra,
photometry and morphology (Eisenstein et al.\ 2001), whereas the MAIN
sample is more hetrogeneous with the general requirement, set by us,
that EW$_{{\rm H}\alpha}<1.5$\,\AA\ (but see Paper I). Even though the
selection is not completely uniform, the resulting sample of E/S0 lens
systems is indistinguishable in its photometric and scaling-relation
properties from non-lens samples (see Paper II).  Nonetheless, we
remain cautious of potential selection effects.

Of the 28 systems observed as of 2005 March 31 -- the cutoff date for
this first series of papers -- we confirmed 19 as unambiguous lens
systems. Of the remaining 9 systems, six show some hint of a
counter-image near the galaxy center, but in all these cases these are
too faint to be confirmed as true lens systems with the present data
(Paper I), while three lack any visible lensed images\footnote{Lensed
images can still be extended and below the noise level of the shallow
Snapshot images.} or are magnfied, but singly imaged, galaxies inside
the SDSS fiber aperture.

The galaxy-subtracted HST--ACS F435W and F814W images of each of the
observed SLACS lens candidates are presented in Paper I. In general,
the F814W images have better signal-to-noise compared to the F435W
images and thus serve as the primary constraint on the lens
models. The F435W images are only used to further validate the lensed
nature of the multiple images (e.g.\ based on similar colors and
structure). 

In this paper, four additional lens systems are removed to construct a
clean sample of ``isolated'' early-type galaxies\footnote{
Because massive galaxies preferentially occur in over-dense regions,
no truly isolated early-type galaxies exist. We therefore only discard
those systems where either the observations or the lens model can
significantly be affected by other nearby massive galaxies. In
general, the latter means two similar galaxies within $\sim$4 Einstein
radii from each other (Kochanek \& Apostolakis\ 1988). Hence,
``isolated'' implies non-interacting and lensing that is dominated by
a single massive galaxy inside the SDSS fibre and several Einstein
radii. This constraint is rather weak, affecting only $\sim$10\% of
the systems.}: one system is a bulge-dominated spiral galaxy
(SDSS\,J1251$-$021), two systems have {\sl two} dominant lens galaxies
inside the lensed images (SDSS\,J1618+439 and SDSS\,J1718+644) and one
system has a nearby perturbing companion that also contributes
significantly to the light inside the SDSS spectral fiber
(SDSS\,J1205+492). The latter makes its stellar dispersion measurement
unreliable. All fifteen remaining systems (see
Table~\ref{tab:results}) are genuine massive early-type lens galaxies,
with a redshift range of $z=0.06 - 0.33$ and an average stellar
velocity dispersion of $\langle \sigma_{\rm ap} \rangle = 263 \pm 11
$\,km\,s$^{-1}$ (rms of 44\,km\,s$^{-1}$) inside the SDSS
spectroscopic aperture. All conclusions in this paper are based on
this sample of 15 galaxies, except for those presented in \S\,4 and
\S\,5.

\subsection{The SIE Mass Model}

The purposes of the lens models are three-fold: (i) Confirm that the
systems are genuine gravitational lenses and that they can be
explained through a simple strong lens model, (ii) accurately
determine the mass enclosed by the lensed images, and (iii) quantify
the alignment between the stellar and total mass distribution.

To determine the mass enclosed by the lensed images, we follow the
procedure described previously in e.g.\ Koopmans \& Treu (2003) and
Treu \& Koopmans (2002, 2004). First, we determine the ``best-fit''
elliptical lens mass model and for that, derive the mass (M$_{\rm
Einst}$) enclosed by the outer (tangential) critical curve. Second, we
determine the associated circularly-symmetric mass model -- having the
same radial density profile -- that encloses the same mass inside its
critical curve at radius R$_{\rm Einst}$, which we call the Einstein
radius.

We use the parametric {\sl Singular Isothermal Ellipsoid} mass model
(SIE; Kormann et al. 1994) to describe the projected mass
distribution (i.e.\ convergence) of the lens galaxies (appropriately
translated and rotated):
\begin{equation}
  \kappa(x,y) = \frac{b_{\rm SIE} \sqrt{q_{\rm SIE}}}{2 \sqrt{q_{\rm SIE}^2
  x^2 + y^2}},
\end{equation}
with $q_{\rm SIE}=(b/a)_{\kappa}$ being the axial ratio of constant
elliptical surface density contours. Note that the mass enclosed by
the elliptical critical curves, using the above normalization, is
independent of $q_{\rm SIE}$ (Kormann et al. 1994). The definition of
the enclosed mass (M$_{\rm Einst}$) and the Einstein radius (R$_{\rm
Einst}$) then correspond to those for a classical Singular Isothermal
Sphere (SIS with $q$=1; e.g.\ Binney \& Tremaine 1987) and can be
associated with a velocity dispersion through $b_{\rm SIE} = 4\pi
(\sigma_{\rm SIE}/c)^2 D_{\rm d} D_{\rm ds}/D_{\rm s}$ (see Schneider
et al.\ 1992). This velocity dispersion should {\sl not} be confused
with that of the stellar component embedded in an overall isothermal
(i.e.\ $\rho_{\rm tot} \propto r^{-2}$) mass distribution (e.g.\
Kochanek 1994). The two quantities can differ, depending on the
precise distribution of the stars, their orbital structure inside the
overall potential, and the aperture within which the dispersion is
measured (Koopmans 2004).

\subsection{Non-Parametric Source \& Image Reconstructions}

The complexity of many of the extended lensed images (Paper I)
prohibits a simple parameterized description of the source (e.g.\
point images). Their brightness distributions are therefore
reconstructed on a grid of typically $30\times30$ square pixels, with
a pixel-size that depends on scale of the lensed images and their
magnification (typically between $0\farcs01-0\farcs05$). We use the
regularized non-parametric source reconstruction code described in
TK04 and Koopmans (2005) -- based on the non-parametric source
reconstruction method by Warren \& Dye (2003) -- with the parametric
SIE mass model for the lens potential. We emphasize that the choice of
isothermal lens models influences the mass determination within the
critical line at a level of a few percent at the most (e.g. Kochanek
1991). This systematic uncertainty is at present negligible in our
analyses of the logarithmic density slope (see \S~3.4 for a proper
discussion), and to first order our lensing and stellar-dynamical
analyses can be regarded as independent.

We center the mass model on the brightness peak of the lens
galaxy\footnote{In several test-cases we find that the mass centroid
agrees with the brightness peak to within a pixel. Because the precise
position of the mass centroid has negligible effect on the inferred
mass enclosed by the lensed images -- or the other lens properties --
we choose to fix the mass centroid position in order to speed up the
convergence process.} and vary the three remaining model parameters
(i.e.\ lens strength $b_{\rm SIE}$, ellipticity $q_{\rm SIE}$ and
position angle $\theta_{\rm SIE}$). At each optimization step, we
determine the source structure that minimizes the value of the penalty
function $P= \chi^2 + \lambda\,R$, which includes a $\chi^2$ and a
regularization term. The mass-model parameters and $\lambda$ are
varied until $\chi^2/{\rm d.o.f.}$ minimizes to $\approx 1$ (see
Warren \& Dye 2003 or Koopmans 2005 for details).

Because the main objective in this paper is to obtain the mass of the
galaxy enclosed by its Einstein radius -- for subsequent dynamical
analyzes (\S 3) -- neither the precise choice of the source pixel
scale nor the regularization level (i.e.\ the value of $\lambda$) is
found to have a significant impact on the resulting values of R$_{\rm
E}$ and M$_{\rm Einst}$. We therefore postpone a precise analysis of the
structure of the sources to a future publication. The SIE models are
sufficiently accurate to (i) confirm the lensed nature of each system,
(ii) measure the mass enclosed by the lensed images to a few percent
accuracy (see above discussion), and (iii) determine the orientation
of the mass distribution with respect to its stellar distribution.

Figure~\ref{fig:siemodels} shows the observed structure of the lensed
images in the F814W band -- after lens galaxy subtraction (Paper I) --
and the currently best lensed-image reconstruction for each of the
fifteen selected lens systems. None of the models require significant
external shear above a few percent to improve the models. Therefore we
can assume it to be zero for simplicity (see \S\,2.4.1 for more
discussion).

Table~\ref{tab:results} summarizes the best-fit parameters from the
SIE mass modeling. We find the ratios $\langle {\rm R}_{\rm
Einst}/{\rm R}_{\rm ap} \rangle = 0.87 \pm 0.05$ (rms of 0.19) between
the Einstein radius and the SDSS fiber radius of 1\farcs5, and
$\langle {\rm R}_{\rm Einst}/{\rm R}_{\rm e} \rangle = 0.52 \pm 0.04$
(rms of 0.17) between the Einstein radius and the effective radius
(see Paper II), with $\langle {\rm R}_{\rm Einst} \rangle = 4.2 \pm
0.4$\,kpc (rms of 1.6\,kpc). Hence, when discussing the ``inner
regions'' of early-type galaxies, we assume this to be approximately
the inner 4~kpc.

\subsection{Stellar versus Total Mass}

Using the SIE mass models, we can assess how well light (i.e. stellar
mass) traces the total mass density. In hindsight, this correlation is
not surprising, because most of the mass inside R$_{\rm Einst}$ is in fact
stellar (see \S\,3.5). However, a significant misalignment or
difference in ellipticity between the stellar and dark-matter mass
components -- even {\sl outside} the Einstein radius -- would affect
the lens models and show up as differences or increased scatter in the
position-angle difference and ellipticity ratio between stellar and total
mass. Note also that we implicitly assume an isothermal density
profile -- which will be further supported in \S\,3.2 -- which in
principle could affect the determination of the mass ellipticity and
position angle.

\subsubsection{Position-Angle Alignment}

One test of the lens mass models is the position-angle alignment
$\Delta \theta = (\theta_* - \theta_{\rm SIE})$ between the stellar
component and the SIE lens model. The result is shown in
Fig.\,\ref{fig:align}. The average difference is $\langle \Delta
\theta \rangle = 0 \pm 3$~degrees, with an rms spread of
10~degrees. No significant correlation is found between $\Delta
\theta$ and other lens properties. One notices an increase in the rms
of $\Delta \theta $ with increasing $q_{\rm SIE}$, because it becomes
increasingly more difficult to determine both $\theta_*$ and
$\theta_{\rm SIE}$ for $q_{\rm SIE}\rightarrow 1$: the rms for $q_{\rm
SIE}<0.75$ is 3 degrees, whereas it increases to 13 degrees for
$q_{\rm SIE}>0.75$. However, no significant deviations of $\langle
\Delta \theta \rangle$ from zero are found in either bin. 

Assuming that (Keeton, Kochanek \& Seljak\ 1997)
\begin{equation}
  \langle \Delta \theta^2 \rangle^{1/2} \approx \left \langle \left(
  \frac{\sin(2 [ \theta_{\rm SIE}-\theta_{\gamma}])}{
  (\epsilon/3\gamma_{\rm ext})\cos(2 [ \theta_{\rm
  SIE}-\theta_{\gamma}])}\right)^2 \right\rangle^{1/2},
\end{equation}
where $\gamma_{\rm ext}$ is the external shear, $\theta_{\gamma}$ the
shear angle, and $\epsilon = (1-q_{\rm SIE}^2)/(1+q_{\rm SIE}^2)
\approx 0.24$ (see \S \ref{sect:ellipticity}) and that no correlation
between galaxy and external shear orientations exists, we then find
that the rms of 10 degrees in $\Delta \theta$ implies that that the
average shear has an upper limit $\langle \gamma_{\rm ext} \rangle \la
0.035 $. Hence, the alignment between the mass and light position
angle confirms that external shear is very small and can be neglected,
and that the galaxies are effectively isolated in terms of their
gravitational lens properties in the inner $\sim$4\,kpc. (Note that this
does not imply that galaxies {\sl are} isolated, only that the effect
of the field on their lensing properties is small.)

\subsubsection{Ellipticity}\label{sect:ellipticity}

\begin{figure}
\begin{center}
\leavevmode
\vbox{%
\epsfxsize=0.69\hsize
\epsffile{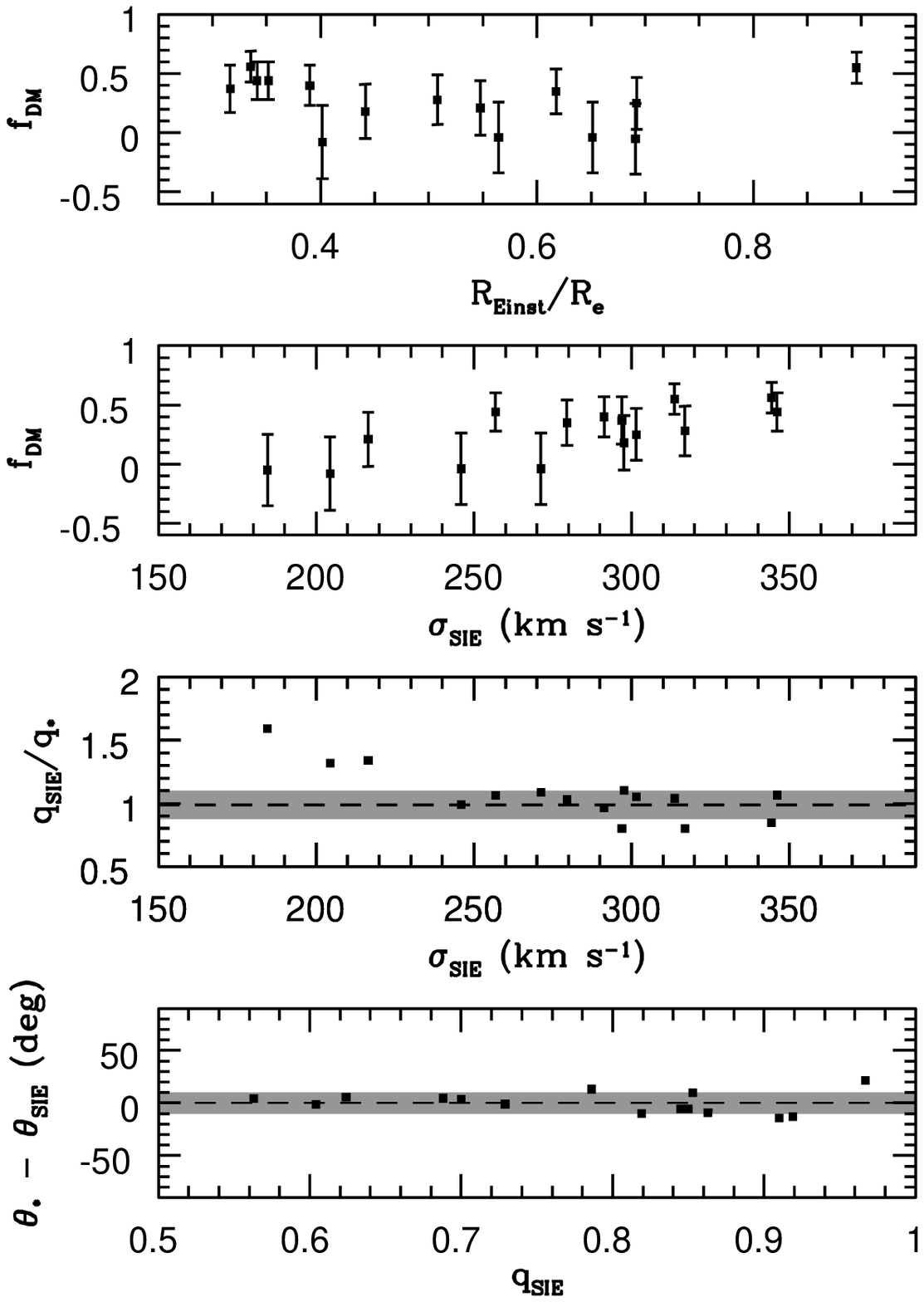}
}
\end{center}
\caption{\label{fig:align} (Upper two panels) Inferred dark-matter mass
fraction inside the Einstein radius, assuming a constant stellar
$M/L_{\rm B}$-ratio as function of E/S0 velocity dispersion (see
text).  (Third panel) Ratio between the ellipticity
(i.e. $q_*=(b/a)_*$) measured from the stellar light (Paper II) and
that determined from the SIE mass model ($q_{\rm SIE}$).  Note the
tight scatter around unity above $\sim 225$~km\,s$^{-1}$ and then the
upturn in $q_{\rm SIE}/q_*$, resulting from more disky S0
systems. (Lower Panel) Difference between the position angle measured
from the stellar light ($\theta_*$; Paper II) and that determined from
the SIE mass model ($\theta_{\rm SIE}$).  Note the increase in rms
dispersion with increasing $q_{\rm SIE}=(b/a)_\Sigma$. The rms spread
for $q_{\rm SIE}<0.75$ is only 3 degrees, whereas for $q_{\rm
SIE}>0.75$ it increases to 13 degrees.}
\end{figure}

A second test is to see how well the elliptical isophotal and
isodensity contours trace each other. Fig.\,\ref{fig:align} also shows
the ratio between the ellipticity of the stellar light (Paper II;
$q_*=(b/a)_*$) and that of the lens mass model ($q_{\rm SIE}$), as a
function of velocity dispersion (i.e.\ approximately mass). Above
$\sigma_{\rm SIE}$ of $\sim$225 km\,s$^{-1}$, the ratio $\langle
q_{\rm SIE}/q_* \rangle = 0.99$ with an rms of 0.11, hence light
traces mass also in ellipticity. Below $\sim$225 km\,s$^{-1}$, the
correlation shows a sudden upturn to $q_{\rm SIE}/q_* \sim$1.6, which
we attribute to the fact that those three galaxies (i.e.\ J0959+042,
J1330-018 and J1420+603) show inclined disky structure and can be
classified as lenticular (S0) galaxies (see Paper I). These results
strongly suggest that the SIE mass model (further supported in \S 3)
quantifies the mass ellipticity to $\sim$10\% accuracy. We can
therefore take the average of $q_{\rm SIE}$ as a good measure of the
projected isodensity ellipticity of early-type galaxies:
$\langle q_{\rm SIE}\rangle$=0.78 with an rms of 0.12. We note that
this ellipticity is that of the stellar plus dark-matter mass
distribution, not that of the dark-matter halo only. Using
\begin{equation}
  \langle q_{\rm SIE} \rangle = 
  \langle [q_3^2 \cos(i)^2 + \sin(i)^2]^{1/2} \rangle,
\end{equation}
assuming that mass is stratified on oblate constant density ellipsoids
and lens-galaxies are randomly oriented (i.e.\ $P(i)\propto \sin(i)$),
this implies an axis ratio in density of $\langle q_{3} \rangle =
(c/a)_{\rho} = 0.66$ with an error of about 0.2.

As an additional check, we assess whether the ellipticity
distributions of the SLACS E/S0 lens-galaxy sample and early-type
galaxies could possibly be different, which would suggest a possible
selection bias. For the 15 SLACS E/S0 lens galaxies, we find $\langle
q_* \rangle = 0.74$ with an rms of 0.13, in excellent agreement with
nearby E/S0 galaxies (e.g.\ Lambas et al.\ 1992; Odewahn et al.\ 1997)
which peak between 0.7--0.8. We conclude also that the ellipticities
of SLACS lens galaxies are similar to those of nearby early-type
galaxies.

\subsubsection{Does Light follow Mass?}

We conclude: (i) The small position-angle difference between the
stellar and total mass implies that dark matter is aligned with the
stellar component on scales $\la 4$~kpc and probably also beyond.
Even though stellar mass dominates in this region (see \S\,3.5), a
misalignment of stellar and dark matter, even beyond the Einstein
radius, can cause an apparent ``external'' shear (Keeton, Kochanek \&
Seljak 1997), which is not observed in our sample. (ii) Significant
external shear due to nearby galaxies or a misalignment of the outer
dark-matter halo with the inner stellar-dominated region is not
required in any of the lens model. Significant external shear would in
general cause a spread in $\Delta \theta$ if not accounted for in the
models. (iii) The isophotal and isodensity contours of massive
elliptical galaxies ($\ga$225 km\,s$^{-1}$) seem to follow each other
well in their inner regions, whereas the lower velocity dispersion
lenticular galaxies have a much rounder mass than light distribution.

\subsubsection{The surface brightness bias of SLACS lenses}

In Papers I \& II, we discussed a bias in favor of more concentrated
light-distributions for lens galaxies compared to their parent
population with equivalent $\sigma_{\rm ap}$.  Because the S/N limit
imposed on the velocity-dispersion measurements of lens-galaxy
candidates was also imposed on the parent sample, and their
distribution in S/N can not be distinguished according to a K-S test,
we concluded in Paper II that a bias in S/N due to the finite fiber
size is not the underlying cause.

Here we propose another bias that might cause part of this effect.
The bias arises because the parent population is chosen to have the
same value of $\sigma_{\rm ap}$ as that of the lens galaxy, but not
necessarily the same mass, i.e.\ lens cross-section. Suppose we have
two galaxies with identical isothermal density profiles and identical
velocity dispersions, $\sigma_{\rm ap}$, inside the fiber aperture,
but different masses (with M$_1$$<$M$_2$). Because the first galaxy is
less massive, the stellar component must be more extended to maintain
a similar velocity dispersion inside the aperture. We find that for
the range R$_{\rm e}/{\rm R}_{\rm ap}\approx 0.8 - 3.2$ (see
Table~\ref{tab:results}) the decrease in mass, and therefore lens
cross-section, is about 5\% for a fixed $\sigma_{\rm ap}$. Whether
this bias can partly explain the observed bias, or whether other
mechanisms are also important, is not clear and we leave a full
analysis to a future paper.

\begin{figure*}[t]
\begin{center}
\leavevmode
\hbox{%
\epsfxsize=0.99\hsize
\epsffile{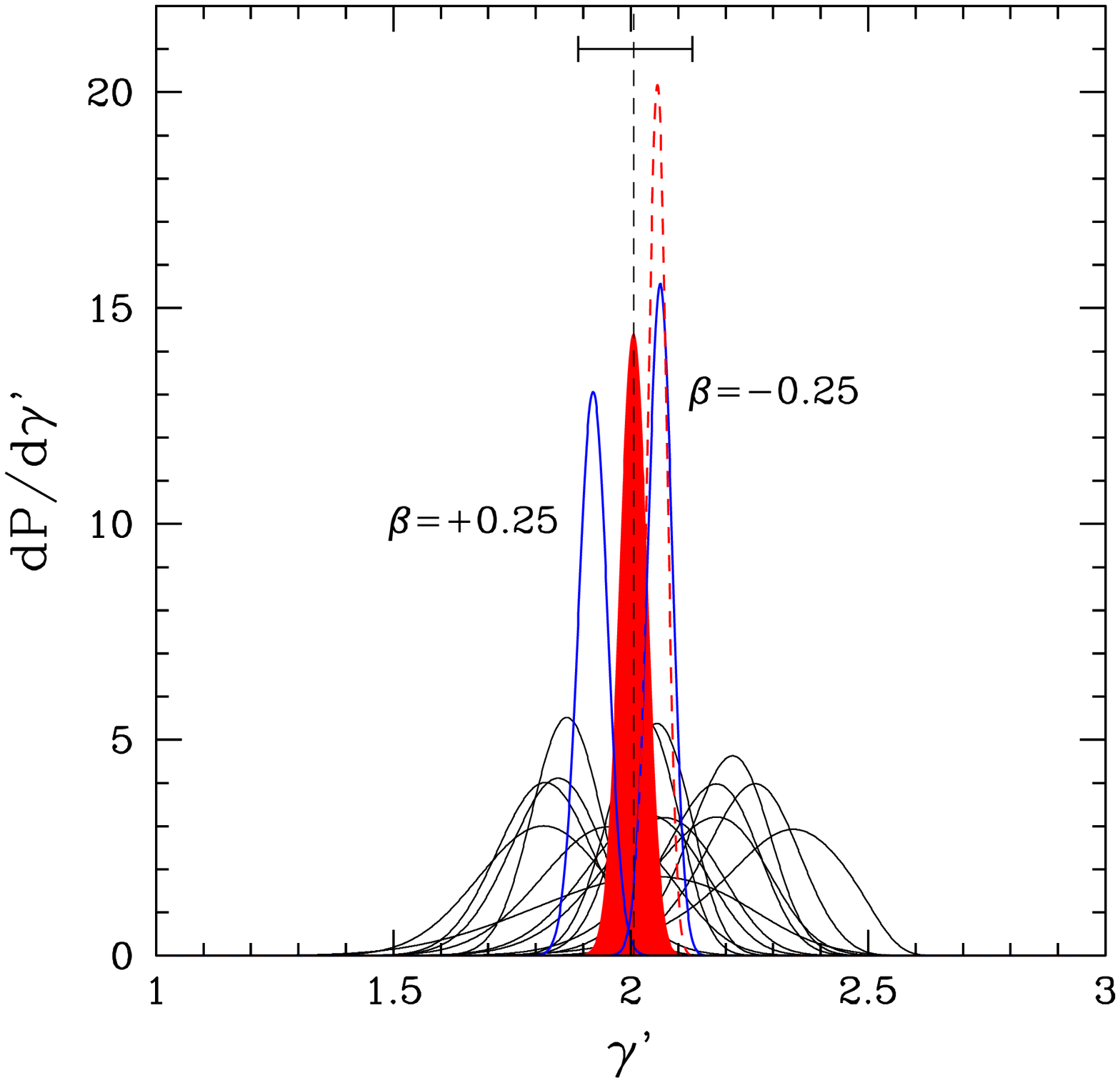}}
\end{center}
\caption{\label{fig:logslope} Posterior probability distribution
functions of the logarithmic total density slope ($\gamma'$; see
text). The shaded region (red) indicates the joint probability for
$\gamma'$, assuming isotropic stellar orbits and a Hernquist~(1990)
luminosity density profile. The thin solid curves refer to the 15
individual lens systems. The dashed (red) curve assumes a Jaffe~(1983)
luminosity density profile, leading to a several percent increase in
the maximum-likelihood value of $\gamma'$. The two solid (blue) 
curves, indicated by $\beta = \pm\,0.25$, show the probability
functions for radially and tangentially anisotropic stellar orbits
respectively (assuming a Hernquist profile for the stellar
component). The horizontal bar indicates the 1\,$\sigma$ intrinsic
spread in $\gamma'$, corrected for the spread due to measurement
errors on the stellar velocity dispersions.}
\end{figure*}

\section{Joint Lensing \& Dynamical Analysis}

In this section, we combine the projected 2D mass-measurement (M$_{\rm
E}$; see \S\,2) with the stellar velocity dispersion measurement from
SDSS spectroscopy ($\sigma_{\rm ap}$) and the surface brightness
distribution from the HST images (see Papers I \& II), to determine
the 3D logarithmic density slope inside the Einstein radius of each
galaxy.

\subsection{Spherical Jeans Modeling}

We model each early-type galaxy as a spherical system, previously
discussed in Koopmans \& Treu (2003) and Treu \& Koopmans (2002,
2004). The modeling is done according to a number of steps and
assumptions:

\begin{itemize}

\item The stellar plus dark-matter mass distribution of each of the
  lens galaxies is modeled as
  \begin{equation}
    \rho_{\rm tot}(r) = \rho_0 \left(r/r_0\right)^{-\gamma'},
  \end{equation}
  where $\rho_0$ can be uniquely determined from the projected mass
  M$(\le {\rm R}_{\rm Einst})$$\equiv$M$_{\rm Einst}$ and $r_0$ can be
  set arbitrarily. In \S 3.5 we discuss in more detail why we make
  this assumption for the familiy of total-density profiles. The only
  remaining free parameter in the density distribution is therefore
  the logarithmic density slope $\gamma'$ [Note that $\gamma' =
  -d\log(\rho_{\rm tot})/ d\log(r)$].  As discussed in TK04, the
  results are very insensitive to a cutoff at large radii (i.e.\
  beyond several effective radii) in the dynamical analysis.

\item The stellar component is treated as a massless tracer (i.e.\
  ${\rm M}_*/{\rm L} \rightarrow 0$) in the gravitational potential of the
  total density profile. We assume a stellar density
  \begin{equation}
    \rho_*(r) = \frac{(3-\gamma_*) {\rm M}_* r_*}{4 \pi
    r^{\gamma_*} (r+r_*)^{(4-\gamma_*)}},
  \end{equation}
  where ${\rm M}_* = {\rm L_{\rm B}} \times ({\rm M}_*/{\rm L_{\rm
  B}})$ is the total stellar mass (assumed to be zero in the limit),
  $r_*$ is a break-radius and $\gamma_*$ in the inner logarithmic
  stellar-density slope. For the Hernquist (1990) profile $\gamma_*=1$
  and $r_* = {\rm R}_{\rm e}/1.8153$, such that half of the projected
  light is inside ${\rm R}_{\rm e}$. The projected Hernquist profile
  closely resembles an $R^{1/4}$ profile with which we determined the
  effective radii of each lens galaxy (see Paper II). In case of the
  Jaffe (1983) profile, $\gamma_*=2$ and $r_* = {\rm R}_{\rm
  e}/0.7447$. The two profiles delineate a range of possible models,
  bracketing the observed range of galaxy profiles, useful to test for
  potential systematics.

\item Given the total density (i.e.\ gravitational potential) and the
  luminosity density, we solve the spherical Jeans equations (see
  Binney \& Tremaine 1987), to determine the line-of-sight stellar
  velocity dispersion as function of radius. The calculations are done
  assuming different (constant) values for the velocity anisotropy of
  the stellar orbits (see e.g.\ Gerhard et al.\ 2001) with $\beta
  \equiv 1-{\langle v_\theta^2 \rangle}/{\langle v_r^2
  \rangle}$. Tangential anisotropy has $\beta<0$, whereas radial
  anisotropy has $\beta>0$.

\item Both seeing and aperture effects are accounted for in the
  dynamical models. The observed stellar velocity dispersion is a
  luminosity-weighted average dispersion inside the SDSS fiber
  aperture. We assume Gaussian seeing with $\langle{\rm
  FWHM}\rangle$=1.5\,arcsec for the SDSS spectroscopic observations,
  although the exact value is almost irrelevant, given the 
  three-arcsec diameter spectroscopic fiber aperture.

\item The probability density of $\gamma'$ is then given by
  \begin{equation}
    \frac{d P}{d \gamma'} \propto e^{-\chi^2 /2},
  \end{equation}
  with $\chi^2 = [(\sigma_{\rm ap}- \sigma_{\rm mod})/\delta\sigma_{\rm
  a}]^2$ and $\delta\sigma_{\rm ap}$ is the 1\,$\sigma$ error on the
  aperture velocity dispersion measured from the SDSS spectra. The
  integrated probability density function is normalized to unity.

\end{itemize}

\subsection{The Logarithmic Density Slope}

Following the procedure described in \S\,3.1, we determine the
probability density functions for $\gamma'$ for each of the fifteen
early-type galaxies, assuming a Hernquist luminosity density profile
with $\beta=0$.  The results are shown in Fig.\,\ref{fig:logslope}
(thin black solid curves) and summarized in Table~\ref{tab:results}.

Also shown is the joint probability (red shaded area), $P_{\rm joint}
\propto \Pi_i\, ({d P_i}/{d \gamma'})$, from which we determine an
average logarithmic density slope of
$$
\langle \gamma' \rangle = 2.01^{+0.02}_{-0.03}~~~{\rm (68\% ~C.L.)}
$$ 
for the ensemble of galaxies. The fractional spread in
$\gamma'$ of $\sim$10\% is partly due to the measurement error $\delta
\sigma_{\rm ap}$. To determine the true intrinsic spread around
$\langle \gamma' \rangle$ in the ensemble of systems, we approximate
the likelihood function of $\gamma'_i$ for each system by a Gaussian
with a 1\,$\sigma$ error of $\delta\gamma'_{i}$. Second, we assume
that the slope $\gamma'_i$ of each system is drawn from a underlying
Gaussian distribution around $\langle \gamma' \rangle$ with an
intrinsic 1\,$\sigma$ spread of~$\sigma_{\gamma'}$.

The maximum-likelihood solution for $\sigma_{\gamma'}$ (ignoring the
much smaller error on $\langle \gamma' \rangle$) is then found from:
\begin{equation}
  \sum_i \left[\frac {(\gamma'_i-\langle \gamma' \rangle)^2 -
    \sigma_{\gamma'}^2 - \delta{\gamma'_{i}}^{2}}{(\sigma_{\gamma'}^2 +
    \delta{\gamma'_{i}}^{2})^2} \right] = 0.
\end{equation}
The solution of this equation is $\sigma_{\gamma'}=0.12$ for the
sample of fifteen early-type galaxies. This is a very small {\sl
intrinsic} spread of only 6\% around the average value, considering
that many effects have not yet been accounted for. 

\subsubsection{Implications for H$_0$ from lensing}

We comment that the small intrinsic scatter in $\gamma'$ of around 6\%
suggests that these low-redshift ($z \la 0.3$) massive early-type
galaxies could be ideal for measuring H$_0$ from
time-delays\footnote{Note that this requires a variable source. In
case of an active galactic nucleus source, however, the lensed images
often outshine the lens galaxy, making a joint lensing and dynamical
analysis more difficult. However, SNe in star-forming lensed sources
could provide very accurate time-delays (Holz 2001; Bolton \& Burles\
2003; Moustakas et al.\ in prep.).}, with an expected rms scatter of
around 12\% between systems if they are assumed to be perfectly
isothermal (TK04). At higher redshifts -- where both external shear
and convergence from the group and/or large-scale-structure
environment of the lens galaxies become more important -- the same
assumption can lead to a larger (i.e.\ around 30\%) systematic scatter
in H$_0$ (see TK04 for a full discussion). We note also that we
measure the average density slope {\sl inside} R$_{\rm Einst}$,
whereas the time-delay depends on the local density slope {\sl inside
the annulus} between the lensed images (Kochanek 2002), which could
have a different value and most likely a larger scatter (because it is
not averaged).

\subsection{Random Errors on the Density Slope}

The assumption of an isothermal lens mass model (see \S\,2.1)
systematically affects the mass determination of the lens galaxy
inside its Einstein radius at most at the few-percent level (see
Kochanek 1991 and \S\,3.4). In addition, the mass determination has a
random error, which we expect to be very small because of the high
S/N--ratio data that we use to fit the models\footnote{The fractional
random error on the mass is $\delta_{\rm M_{\rm Einst}} \approx 2\cdot
\delta_{\Delta \theta}$ for the SIE mass model. Because the error on
the image separation $\Delta \theta \approx 2 b_{\rm SIE}$ is
typically the width of the lensed arcs divided by twice the S/N-ratio,
we expect the random error on ${\rm M}_{\rm Einst}$ to be less than a
few percent. In the low-S/N case of SDSS J1402+634 (Bolton et al.\
2005), for example, we found a random error of $\sim$3\%. We expect
the other SLACS systems to have much smaller errors because of their
typically much higher S/N-ratio images.}. The alignment of mass and
light (\S\,2.4) also suggests that the remaining degeneracies in the
mass model are too small to lead to a biased mass estimate [on
average].

In the rest of this section we show that residual errors on the
lensing-based mass determination within the critical line are
negligible with respect to the measurement errors on the stellar
velocity dispersion.  In the spherically symmetric case with power-law
dependencies for the luminosity density and total density, one can
show (Koopmans 2004) that the fractional error
$\delta_{\gamma'}\equiv\delta\gamma'/\gamma'\ll 1$ is related to those
on the mass (${\rm M}_{\rm Einst}$) and the measured stellar velocity
dispersion ($\sigma_{\rm ap}$) by
\begin{eqnarray}\label{eq:err1}
  \delta_\sigma(\le {\rm R}_{\rm ap}) &=& \frac{1}{2}
  \delta_{{\rm M}_{\rm Einst}} + \frac{1}{2}\left(\frac{\partial
  \log f}{\partial \log \gamma'} - \gamma' \, \log\left[\frac{{\rm R}_{\rm
  ap}}{{\rm R}_{\rm Einst}}\right]\right)\cdot \delta_{\gamma'} \nonumber\\
  &\equiv& \frac{1}{2} \left( \delta_{{\rm M}_{\rm Einst}} + \alpha_g
   \cdot \delta_{\gamma'}\right),
\end{eqnarray}
from which one finds (assuming independent errors)
\begin{equation}\label{eq:relerr}
  \left\langle \delta_{\gamma'}^2\right\rangle = \alpha_{\rm g}^{-2}
  \left\{\left\langle \delta_{{\rm M}_{\rm Einst}}^2\right\rangle + 4
  \left\langle \delta_{\sigma}^2\right\rangle \right\}.
\end{equation}
Here $\alpha_{\rm g}$ is typically of order a few and the function $f$
depends on the logarithmic slopes of the total and luminosity density
profiles and $\beta$ (see Koopmans 2004 for its full expression in
terms of gamma-functions). The pre-factor of four and the typical
fractional errors on $\delta_{\sigma}\sim 0.05$ from SDSS spectroscopy
(see Table~\ref{tab:results}), implies that $\delta_{{\rm M}_{\rm
Einst}}$ can be neglected given the current kinematic data
quality. The fractional error on the logarithmic density slope is
therefore, to first order, equal to the fractional error on the
measured stellar velocity dispersion. Even though the Hernquist and
Jaffe luminosity density functions follow a broken power-law
(\S\,3.1), this relation holds in our joint lensing and dynamical
analysis with a fractional spread in density slopes $\gamma'$ very
close to that in stellar velocity dispersions (c.f. Paper II and
TK04).

\subsection{Systematic Uncertainties on the Density Slope}

The dominant systematic uncertainties in the current analysis are
probably the unknown stellar velocity anisotropy, the assumption of
spherical symmetry in the dynamical models, deviations of the inner
luminosity density profile from the assumed Hernquist luminosity
density profile and the possible contribution to the mass inside the
Einstein radius by the surrounding field galaxies.

To assess some of these uncertainties, we redo our analysis for a change of
anisotropy parameters $\Delta\beta=[-0.25, +0.25]$ (with a Hernquist
profile).  The resulting change in the average value of $\gamma'$ is
relatively small $\Delta\gamma'=[+0.05,-0.09]$. Similarly, if we assume
a Jaffe (1983) luminosity density profile, we find $\Delta
\gamma'=+0.05$ (assuming $\beta=0$).

Because of their comparable scales, the small value of
$\sigma_{\gamma'}$ could partly be due to some remaining systematic
effects. We note that the ensemble could also be more radially
anisotropic (e.g.\ Gerhard et al. 2001) and/or have a luminosity
density cusp steeper than Hernquist, but the above analysis shows
these systematic shifts in $\gamma'$ to be $\la 5\%$ for reasonable
assumptions. We note also that the small intrinsic spread in
$\gamma'$, in principle, allows us to set an upper limit on the
average anisotropy of their velocity ellipsoid, using the tensor
virial theorem (Binney\ 1978; Sandy~Faber and Chris~Kochanek, private
communications). This could potentially lead to a correlation between
the stellar mass ellipticity and its velocity dispersion and therefore
with the inferred density slope. To test this, we plot $\gamma'$
against $q_*$ in Fig.\ref{fig:plotcorr}: no significant correction is
found and the effect must therefore be small. We defer a more thorough
analysis to a future publication, that will make additional use of
more detailed kinematic data obtained from IFS observations of several
of the SLACS lens galaxies and two-integral dynamical models.

If the field around the lens galaxies contributes significantly to the
enclosed mass (i.e.\ to the convergence inside the Einstein radius),
it biases $\gamma'$ to lower values, if not accounted for (see e.g.\
TK04 for a discussion). There are several reasons why we believe this
contribution to be relative small for the SLACS lens systems. First,
we found that each systems can be modeled as a SIE without requiring
significant external shear. In general the strength of the shear
equals roughly the convergence of the field (if dominated by only a
few systems). Second, because SLACS lens systems are at relatively low
redshifts compared to lens systems known to date (typically at $z_{\rm
l}\sim0.6$), the angular distance between the lens galaxies and their
nearest neighbors (in units of the Einstein radius) is larger than at
higher redshift.  Consequently, the influence of the field on the lens
system is lower by a least a factor of a few compared to high-$z$
systems (e.g.\ Paper II). In addition, because the external convergence
$\kappa_{\rm ext}$ lowers the mass fractionally by $\sim \delta_{\rm
M_{\rm Einst}}$, we find from Equation~\ref{eq:relerr} that
$\delta_{\gamma'} \sim \kappa_{\rm ext}/\alpha_{\rm g}$. Hence, even a
high external shear or convergence of 0.1 -- easily detectable by the
lens models -- would affect $\gamma'$ at most at the $\sim$5\% level
(for typical $\alpha_{\rm g}\sim 2$). On average, however, the
expected external convergence is only a few percent (e.g.\ Fassnacht
\& Lubin 2002; Keeton \& Zabludoff 2004; Dalal \& Watson 2004),
reducing its influence to less than a few percent.

\begin{figure*}[t]
\begin{center}
\leavevmode
\hbox{%
\epsfxsize=0.99\hsize
\epsffile{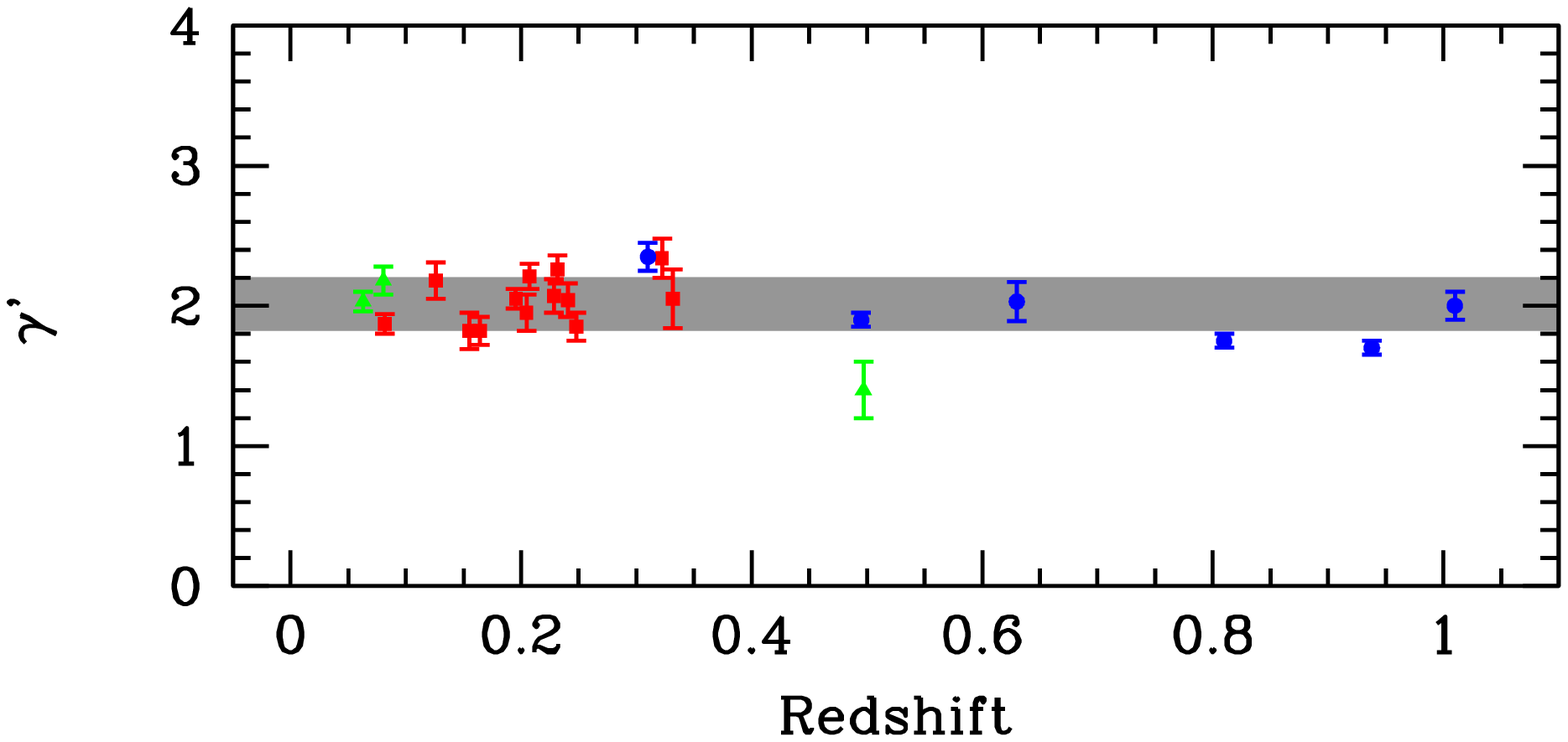}}
\end{center}
\figcaption{\label{fig:gammz}The logarithmic density slopes ($\gamma'
= - d\log \rho_{\rm tot}/d\log r$) of field early-type lens galaxies,
plotted against redshift. The (grey) box indicates the rms spread
(0.19; partly due to measurement errors) around the straight average
of all SLACS/LSD systems with $\sigma_{\rm ap}\ge 200$~km\,s$^{-1}$
(2.01; the same as from the complete SLACS sample). The red solid
squares are from the SLACS Survey, and the blue solid circles from the
LSD Survey plus two additional systems (see text). The green solid
triangles are SLACS/LSD systems with $\sigma_{\rm ap}<
200$~km\,s$^{-1}$.}
\end{figure*}

Another systematic uncertainty could stem from the determination of
$\sigma_{\rm ap}$ from the SDSS spectrum.  If SDSS velocity
dispersions were systematically biased, this would skew $\gamma'$ in
one or the other direction, although there is no reason to assume this
bias exists. Moreover, Bernardi et al.\ (2003) estimate systematic
uncertainties on the measured velocity dispersions of $<$3\%. The
ongoing IFS observations, discussed above, will allows us to
rigorously test this.

\subsubsection{The assumption of a power-law density profile}

Finally, the most serious assumption is the shape of the
density-profile itself (\S 3.1), i.e. a power-law. This assumption can
be tested, however. If either the density profiles of lens galaxies
are different from a power-law, but have the same shape for each
galaxy (scaled to a common scale), or, if they are different from a
power-law {\sl and} different between lens galaxies, in both cases one
expects the inferred (average) logarithmic density slope inside
R$_{\rm Einst}$ to change with the ratio (R$_{\rm Einst}$/R$_{\rm
e}$).  For example, if the profile is a broken power-law with a change
in slope inside R$_{\rm Einst}$, one expects $\gamma'$ to change
depending on where the change in slope occurs with respect to the
effective radius. One would find some ``average'' slope weighed by
luminosity and kinematic profile, and expect this to change as function
of (R$_{\rm Einst}$/R$_{\rm e}$), because R$_{\rm Einst}$ depends mostly
on the relative distances of the lens and the source and is not a
physical scale of the lens galaxy itself. The absence of any clear
systematic correlation between $\gamma'$ and this ratio (see Fig.5),
however, shows that this is not the case. The small deviations of
$\gamma'$ from 2.0 further support this (\S 3.2). We conclude that the
assumption of a single power-law shape for the total density profile 
is valid at the level warranted by the current data.

\subsection{Dark Matter Inside the Einstein Radius}

The spatially resolved kinematic profile and the high-quality data of
{\sl HST} allowed TK04 to do separate analyses of the luminous and
dark matter in individual early-type lens galaxies. The larger SDSS
fiber aperture, the absence of spatially resolved information, and the
higher stellar mass fraction inside R$_{\rm Einst}$ prevent us from
performing a similarly precise analysis. This will require higher
spatial resolution kinematic data (e.g.\ with integral-field
spectroscopy) and is left for future work.

Despite this limitation, we can still infer an average dark-matter
mass fraction inside the Einstein radii of the ensemble of systems,
keeping some caveats in mind (see discussion below).  To do this, we
first solve the spherical Jeans equation for two-component mass models,
assuming a Hernquist luminosity density profile scaled by a stellar
mass-to-light ratio, plus a dark-matter density component with density
profile $\rho_{\rm DM}\propto r^{-\gamma}$, and assuming that $\beta =
0$ (see Koopmans \& Treu 2003 and Treu \& Koopmans 2002, 2004 for more
details). The sum of both mass components must be M$_{\rm Einst}$
inside the Einstein radius. This leads to a likelihood grid as
function of stellar $M_*/L_{\rm B}$ and dark-matter density slope
$\gamma$ (see e.g.\ TK04). Second, a Gaussian prior is set on the
value of $M_*/L_{\rm B}$, assuming an average local restframe B-band
stellar mass-to-light ratio of 7.3~${\rm M}_\odot/{\rm L}_{\odot,\rm
B}$ with a 1--$\sigma$ of 2.1~${\rm M}_\odot/{\rm L}_{\odot,\rm B}$
(e.g.\ Gerhard et al.\ 2001; see also TK04) and correcting this for
the average passive evolution of $d\log(M/L_{\rm B})/dz = -0.69 \pm
0.08$ found from the sample in Paper II, which brightens galaxies with
increasing redshift. The luminosity corrections are small ($\la
20\%$), however, for SLACS galaxies at redshifts below 0.33. Finally,
we marginalize the resulting probability distribution (including the
mass-to-light ratio prior) over $\gamma$ to obtain the likelihood
function of the dark-matter mass fraction inside R$_{\rm Einst}$:
$f_{\rm DM}=1-f_*$. The stellar mass fraction $f_*$ is given by the
maximum-likelihood value of $M_*/L_{\rm B}$ divided by the maximum
allowed value of $M_*/L_{\rm B}$\footnote{The stellar mass must
satisfy M$_*(\le R_{\rm Einst}) \le {\rm M}_{\rm Einst}$}. The results
are listed in Table~\ref{tab:results}. The straight average of $f_{\rm
DM}$ is found to be $\langle f_{\rm DM} \rangle =0.25 \pm 0.06$ (rms
of 0.22) inside $\langle {\rm R}_{\rm Einst} \rangle = 4.2 \pm
0.4$\,kpc (rms of 1.6\,kpc) with a large range between about 0\% to
60\%.

We note that none of the stellar mass-to-light ratios significantly
exceeds the maximum set by the inequality M$_*(\le R_{\rm Einst}) \le
{\rm M}_{\rm Einst}$, which can be regarded as an additional (although
weak) check on the validity of our assumptions. Finally, we
investigate whether the dark-matter fraction ($f_{\rm DM}=1-f_*$)
inside the Einstein radius correlates with other lens properties. In
Fig.2, we plot $f_{\rm DM}$ against $\sigma_{\rm SIE}$ and do indeed
find a correlation (correlation coefficient $r=0.74$).  This
correlation is predominantly due to the low inferred dark-matter mass
fraction in the low-mass (i.e.\ low dispersion) galaxies, whereas the
higher mass systems seem to have $f_{\rm DM}\approx 0.4-0.6$. This
correlation can either be true, i.e.\ low-mass galaxies have less dark
matter in their inner regions (Napolitano et al.\ 2005), or be a
result of the break-down of one of our assumptions.

We think that the latter is the most likely explanation, because the
three lowest-mass systems with $f_{\rm DM}\approx 0$ (all S0 galaxies)
have values of $q_{\rm SIE}/q_* \gg 1$, suggesting that dark matter
contributes significantly to their inner regions similar to spiral
galaxies. In fact, the assumption of a constant stellar $M/L_{\rm B}$
ratio, independent of $\sigma_{\rm SIE}$, is probably not entrirely
correct. Within the ``down-sizing'' scenario which leads to a tilt in
the Fundamental Plane (see e.g.\ Paper II), a lower stellar $M/L_{\rm
B}$ ratio is expected for S0 galaxies. This leads to a lower inferred
dark-matter mass fraction inside the Einstein radius. In addition,
we find no strong trend of dark-matter mass fraction with the ratio
of Einstein radius over effective radius (Fig.2).

We therefore conclude that the inferred values of $f_{\rm DM}$ only
give a general indication\footnote{Note that the dark-matter mass
fraction is not given within a physical radius, such as the effective
radius, because it requires a more proper density model for stellar
and dark matter which is not possible with the current data.}. The
other results in this paper, however, do not depend on the stellar
$M/L_{\rm B}$ assumption. As mentioned above, we are obtaining higher
spatial resolution kinematic data to improve upon this without
assuming a stellar mass-to-light ratio.

\begin{figure*}[t]
\begin{center}
\leavevmode
\hbox{%
\epsfxsize=0.8\hsize
\epsffile{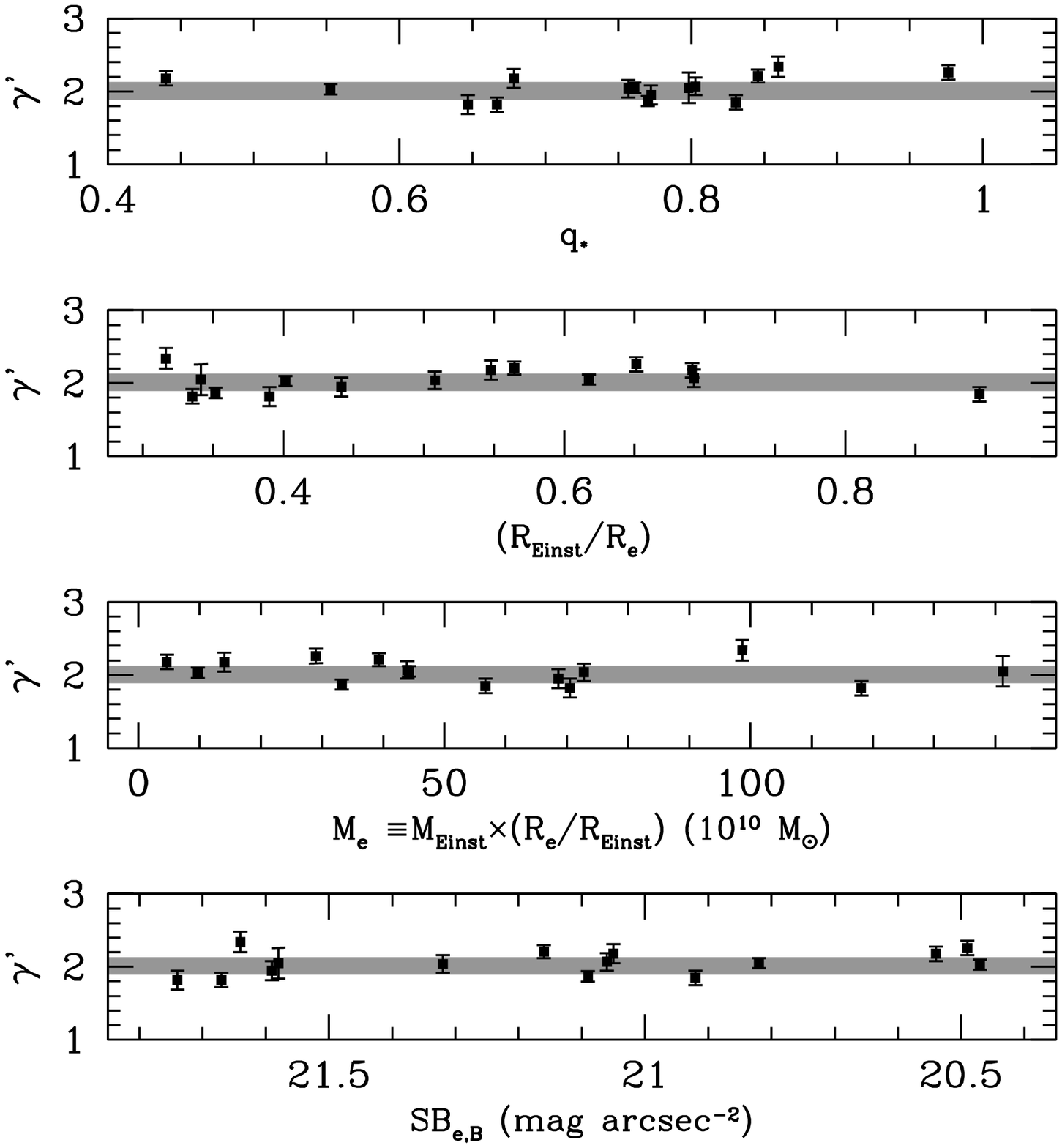}}
\end{center}
\figcaption{\label{fig:plotcorr}The logarithmic density slope of SLACS
lens galaxies as function of stellar ellipticity, the
(normalized) Einstein radius, the projected mass inside the effective
radius and the effective surface brightness, respectively. In none of
these cases, a significant correlation is found. In particular, the
absence of any correlation between $\gamma'$ and SB$_{\rm e,B}$ argues
against more condensed early-type galaxies having a steeper density
profile. Similarly, the absence of correlation with $(R_{\rm
Einst}/R_{\rm e})$ suggests that in the transition region from a
stellar to dark matter dominated density distribution, the logarithmic
density slope is unchanged.}
\end{figure*}

\section{The Evolution of the Inner Density Profile of Early-type Galaxies} 

In combination with results from the Lenses Structure \& Dynamics
(LSD) survey (Koopmans \& Treu 2002, 2003; Treu \& Koopmans 2002,
2004) and from two more lens systems for which stellar velocity
dispersions are measured and analyzed in a homogenous way (Treu \&
Koopmans 2002b; Koopmans et al. 2003) at $z\ga 0.3$, we are now in a
position to measure the evolution in the average logarithmic density
slope of early-type galaxies to $z\approx 1$, if present. Since these
systems have been selected in different ways, we initially limit our
analysis to the most massive early-type galaxies with $\sigma_{\rm
ap}>200$~km\,s$^{-1}$ (i.e.\ $\ga$\,L$_*$). Because the brightness
profile, stellar velocity dispersion, and lens models are easier to
obtain for these systems, we can expect smaller systematic
uncertainties in this sample. Only 3 systems do not meet this
criterion. This cut also simplifies comparisons to massive galaxies in
numerical simulations (e.g.\ Meza et al. 2003; Kawata \& Gibson 2003).

Fig.\,\ref{fig:gammz} shows all systems from the SLACS and LSD
surveys, including PG1115+080 (Treu \& Koopmans 2002) and
B1608+656--G1 (Koopmans et al.\ 2003). The unweighted average value of
$\gamma'$ for the LSD/SLACS systems with $\sigma_{\rm
ap}>200$~km\,s$^{-1}$ is 2.01 (similar to the SLACS sample alone) and
an rms of 0.19 is found.  To measure the evolution of the density
slope, we do an unweighted linear fit\footnote{Because of the relative
small ensemble of systems at higher redshifts and their less well-know
selection effect, the intrinsic spread in $\gamma'$ might be poorly
sampled. A weighted fit would then bias the result to a few systems.}
to the ensemble of systems, finding $\langle \gamma'\rangle (z)=(2.10
\pm 0.07) - \alpha_{\gamma'} \cdot z$ with
$$
  \alpha_{\gamma'}\equiv \frac{d \langle \gamma' \rangle}{d z} =
  0.23 \pm 0.16~~{({\rm 1}\,\sigma)}
$$ below $z \la 1$. This is marginally consistent with {\sl no}
evolution, or with a change in $\langle \gamma' \rangle$ of $\sim
10\%$ in the last $\sim$7~Gyr. If we include those systems with
$\sigma_{\rm ap}<200$~km\,s$^{-1}$ we find $\alpha_{\gamma'} = 0.29
\pm 0.17$. More recently, Hamana et al.\ (2005) analyzed two more
systems, B2045+265 (Fassnacht et al.\ 1999) and HST\,14113+5211
(Fisher et al.\ 1998); inclusion of their results gives
$\alpha_{\gamma'} = 0.34 \pm 0.15$. However, the former lens system
has a disputed source redshift and the latter a massive nearby
cluster, leading us to not select these systems in the LSD
Survey. Despite combining variously-selected lens systems, the results
are robust against changes in the cut in $\sigma_{\rm ap}$ or the
selection of included lens systems. 

To our knowledge, this is the first constraint on the evolution of the
logarithmic density slope in the inner regions of early-type galaxies
to redshifts as high as $z\approx 1$, although we note that a
progenitor bias could play a role here in that we only select those
systems (from the SDSS LRG and MAIN samples) that have not undergone
recent major mergers that resulted in star-formation (i.e.\ large
EW$_{{\rm H}\alpha}$; although major dry mergers could still have
occured).

\subsection{Other Correlations}

To test whether $\gamma'$ correlates with other quantities of
interest, in Fig.\,\ref{fig:plotcorr} we plot $\gamma'$, from only the
SLACS early-type galaxies, against (i) the Einstein radius (in units
of effective radius), (ii) the projected mass inside the effective
radius, and (iii) the effective surface brightness. The first allows
us to assess whether the density slopes change over the region where
dark matter becomes more dominant, the second whether the density
slope depends on galaxy mass and the third whether more concentrated
stellar distribution imply a more concentrated density distribution (see
Papers I \& II). We find in all three cases {\sl no} significant
correlation. We therefore conclude that {\sl at the current level of
significance} the logarithmic density slope appears a constant and
only in combination with the LSD systems a marginal trend with
redshift might be observed.

\section{Summary of Results}


The {\sl Sloan Lens ACS} (SLACS) Survey has provided the largest
uniformly selected sample of massive early-type lens galaxies to date
(Papers I \& II). We used a sub-sample of fifteen early-type
lens galaxies with a redshift range of $z=0.06-0.33$ and an unweighted
average stellar velocity dispersion of $\langle \sigma_{\rm ap}
\rangle = 263 \pm 11$\,km\,s$^{-1}$ for a joint lensing and dynamical
analysis, finding the following results:

\begin{itemize}

\item The average logarithmic density slope of the {\sl total} mass
density of $\langle \gamma' \rangle = 2.01^{+0.02}_{-0.03}$ (68\%
C.L.) assuming a total density profile of $\rho_{\rm tot}\propto
r^{-\gamma'}$ and no anisotropy (i.e.\ $\beta=0$). Systematic
uncertainties (e.g.\ orbital anisotropy and different luminosity
density cusps) are expected to be $\la 5-10\%$ [see \S\,3.2].

\item The {\sl intrinsic} spread in the logarithmic density slope is
at most 6\%, i.e.\ $\sigma_{\gamma'}=0.12$ (the 1\,$\sigma$ of the
assumed Gaussian spread) [see \S\,3.2], after accounting for
measurement errors.

\item The average position-angle difference between the stellar light
component and the total mass component is found to be $\langle \Delta
\theta\rangle = 0 \pm 3$~degrees with 10~degrees rms, setting
an upper limit of $\langle \gamma_{\rm ext}\rangle \la 0.035$ on the
average external shear. [see \S\,2.4].  

\item The ellipticity of the total surface-density is $\langle q_{\rm
SIE}\rangle$=0.78$\pm$0.03 (rms of 0.12) and $\langle q_{\rm SIE}/q_*
\rangle = 0.99 \pm 0.03$ (rms of 0.11) for $\sigma \ga
225$~km\,s$^{-1}$. Assuming an oblate mass distribution and random
orientations, this implies $\langle q_3 \rangle\equiv
\langle(c/a)_\rho\rangle =0.7$ with an error of 0.2 [see \S\,2.4].

\item The unweighted average projected dark-matter mass fraction is
$\langle f_{\rm DM} \rangle =0.25 \pm 0.06$ (rms of 0.22) inside
$\langle {\rm R}_{\rm Einst} \rangle = 4.2 \pm 0.4$\,kpc (rms of
1.6\,kpc) [see \S\,3.5].

\item The evolution of the total density slope for galaxies with
$\sigma_{\rm ap}\ge 200$~km\,s$^{-1}$ ($>$L$_*$), inside half an
effective radius, $\langle \gamma'\rangle (z)=(2.10 \pm 0.07) -
(0.23\pm0.16)\cdot z$ for the range $z=0.08-1.01$ (combined sample
from the SLACS and LSD Surveys) [see \S\,4].

\end{itemize}

\noindent {\it Summarising: Massive early-type galaxies below
$z$\,$\approx$\,1 have remarkably homogeneous inner mass density
profiles, i.e.\ $\rho_{\rm tot} \propto r^{-2}$ (equivalent to a flat
rotation curve for a rotation supported system), and very close
alignment between stellar and total mass. There is no evidence for
significant evolution in the ensemble average logarithmic density
slope of dark plus stellar mass below a redshift of one in their inner
half to one effective radius.}

\section{Discussion \& Open Issues}

Although the isothermal nature of early-type galaxies has previously
been shown through dynamical, X-ray and lensing studies (see\ \S\,1),
our results are the first where the logarithmic density-slope of
individual early-type galaxies have been determined beyond the local
Universe (i.e.\ between $z=0.08$ and 1.01), based on a well-defined
sample of systems from the SLACS Survey (Papers I \& II), complemented
with the most massive systems from the LSD Survey (e.g.\ TK04) at
higher redshift. These combined results provide the first direct
constraint on the evolution of the inner regions of massive early-type
galaxies with cosmic time.

Even though we find no evidence for significant evolution in the inner
regions of massive early-type galaxies, this does {\sl not} require that
early-type galaxies at $z\approx 0$ have the same formation or
assembly epoch as those studied at $z\approx 1$. A similar
``progenitor bias'' as in FP studies might play a role here as well
(e.g.\ van Dokkum \& Franx 2001). This possibility should always be
kept in mind in comparing galaxies at different redshifts.

Because the number-density of massive early-type galaxies does not
appear to have changed by more than a factor of two since
$z$\,$\approx$\,1 (e.g.\ Im et al.\ 2002; Bell et al.\ 2004; Treu et
al.\ 2005a\&b; Juneau et al.\ 2005) and no major evolution of
$\gamma'$ has been found in our combined SLACS plus LSD sample (see
Fig.\,\ref{fig:gammz}), one can conclude that either the time-scale
for their inner regions to relax to isothermality must be very short
(less than a Gyr), if roughly half of the ellipticals in the sample at
$z\la 0.3$ formed at redshifts below one, or that the inner regions of
most early-type observed at $z \la 1$ were already in place at higher
redshifts, consistent with collisionless numerical simulations (e.g.\
Wechsler et al.\ 2002; Zhao et al.\ 2003; Gao et al.\ 2004).

\subsection{Formation Scenarios}

How do these observational results fit into a hierarchical scenario
where massive early-type galaxies form through gas-rich or gas-poor
(i.e.\ ``dry'') mergers? The SLACS and LSD samples provide three core
pieces of information on stellar-population {\sl and} structural
properties of early-type galaxies :

\begin{enumerate}

\item Their inner regions consist of an old stellar population formed
  at $z\gg 1$, with some evidence for secondary infall at lower
  redshifts of at most $\sim$10\% in mass (Paper II).

\item Their inner regions have nearly isothermal density profiles and
  show remarkably little intrinsic spread in their density slopes.

\item Their inner regions show little evolution below $z$$\approx$1 in
  the ensemble average of the density slope.

\end{enumerate}

With these pieces of information, we can examine the likelihood of
different formation scenarios, not only in the context of their
stellar populations e.g. through the FP studies (see Paper II), but
also based on their structural properties and structural evolution
(this paper).  

\subsubsection{Collisional ``Wet'' Mergers}

It has been suggested that massive elliptical form predominantly at
$z<1$ from the mergers of gas-rich disk galaxies (e.g.\ Kauffmann,
Charlot \& White 1996; Kauffmann \& Charlot 1998). Although this
scenario might quickly lead to relaxed galaxies -- after a rapid
star-burst triggered by the gas-shocks and inflows and the subsequent
relaxation of the resulting stellar populations in several dynamical
time-scales (e.g.\ Barnes \& Hernquist 1991, 1992, 1996; Mihos \&
Hernquist 1996) -- the relatively old stellar populations seems to
exclude this scenario as a dominant effect at low redshifts in the
SLACS and LSD early-type galaxies (see Paper II).

Similarly, we conclude that no significant ``secular evolution'' in
the form, e.g., of adiabatic contraction (e.g.\ Blumenthal et al.\
1986; Ryden \& Gunn 1987; Navarro \& Benz 1991; Dubinski 1994;
Jesseit, Naab \& Burkert 2002; Gnedin et al.\ 2004; Kazantzidis et
al.\ 2004) of the inner regions ($\sim$4\,kpc) seems to have occurred
at $z$\,$\la$\,1 in the inner regions of the population of early-type
galaxies that was already in place at $z \approx 1$. This would lead
to a continuous increase in their ensemble average inner density slope
toward lower redshifts. In Treu \& Koopmans (2002), we tentatively
concluded this already, based on the analysis of a single lens system
MG2016+112 at $z$=1.01, with an upper limit on its inner dark-matter
density slope only marginally consistent with that predicted by
numerical simulations (e.g.\ Navarro et al.\ 1996; Moore et al.\
1998) after adiabatic contraction. 

A more gradual infall of gas-rich satellites, leading to
secondary episodes of star formation (e.g.\ Trager et al.\ 2000; Treu
et al.\ 2002) seems limited on average to $\sim$10\% in mass below
redshifts of unity (see also Treu et al.\ 2005a,b). Such an infall
could alter the average structural properties of the population of
early-type galaxies; at present, we can not exclude that the
(marginally) positive value of $\alpha_{\gamma'}= 0.23\pm0.16$ (see \S
4) could be due to a slight change of the inner regions of massive
ellipticals, as a result of secondary gas-infall. However, it does not
appear to be a dominant effect in the structural evolution of most
massive early-type galaxies at low redshifts.

Hence, the observational evidence appears to show that most of the
massive  SLACS and LSD early-type galaxies were already in place at
$z\approx 1$ in terms of their dominant old stellar population {\sl
and} dynamically, although very rapid ``dry'' mergers of several of 
the lower-redshift galaxies can not be fully excluded. 

Beyond redshifts of $z \approx1$ and in disk galaxies (which might
later assemble into massive elliptical galaxies), gas infall and
dissipational processes are most likely very important (e.g.\ Mo, Mao
\& White 1998; Abadi et al. 2003a\&b). The total density profile of
the simulated disk galaxy at $z \approx 4$ in Gnedin et al.\ (2004),
for example, is close to isothermal in the inner 1--4\,kpc, although
steeper inside the inner $\sim$1\,kpc.  Hence, even though these
simulations suggest that adiabatic contraction plays a role in the
formation for disk galaxies at high redshift, it remains unclear how
it could lead to such a tight intrinsic scatter of $\la$6\% in the
logarithmic density slopes around $\gamma' = 2$ of early-type galaxies
(see \S\,3.2), that presumably form from the mergers of these disk
galaxies (e.g.\ Toomre \& Toomre 1972; Gerhard 1981; Negroponte \&
White 1983; Barnes 1988; Hernquist 1992). We note that the formation
of elliptical galaxies from gas-rich mergers at $z\gg 1$, and
associated star-formation and gas-depletion, would be consistent with
their observed old stellar populations.

\subsubsection{Collisionless ``Dry'' Mergers}

The isothermal nature of the inner regions of early-type galaxies,
already at look-back times of $\sim$7\,Gyrs, is often explained by a
very violent assembly of these regions from collisionless matter,
i.e.\ stars and dark matter (e.g.\ Lynden-Bell 1967; van Albada 1982;
Stiavelli \& Bertin 1987). Even though we indeed find isothermal mass
density profiles, this is remarkable given the problems with violent
relaxation models (e.g.\ Arad \& Lynden-Bell 2005; Arad \& Johansson
2005). If we were to consider formation via mergers of collisionless
stellar systems (i.e. ``dry mergers''), as suggested by some authors
(e.g.\ Kochfar \& Burkert 2003; Nipoti et al.\ 2003; Boylan-Kolchin et
al.\ 2004; Bell et al.\ 2005; Naab et al.\ 2006) we would run into an
additional problem. Dehnen (2005) recently showed that the inner cusps
of remnants in collisionless merging can {\sl not} be steeper than the
cusps of any of the collisionless progenitors and most likely also not
more shallow. This invariance seems in agreement with collisionless
numerical simulations (e.g.\ Wechsler et al.\ 2002; Zhao et al.\ 2003;
Gao et al.\ 2004; Kazantzidis et al.\ 2005).  The
isothermal mass density profile of massive elliptical galaxies would
therefore seem to imply that their ``dry'' progenitors must have had
isothermal profiles as well, {\sl ad infinitum}. Clearly this sequence
must break down somewhere, if one wants to reconcile our finding with
the outcome of cosmological numerical simulations that indicate inner
density cusps with logarithmic density slopes around 1.0--1.5 for
collisionless mergers (Navarro, Frenk \& White 1996; Moore et al.\
1998).

\subsection{Hierarchical Dry Merging of Collisionally Collapsed 
Gas-rich Progenitors}

To bring all the different pieces of evidence for and against gas-poor
or gas-rich formation scenarios together and in agreement with the
observations from the SLACS and LSD samples, we consider the following
scenario: 

\noi (1) At a redshift of approximately 1.5 or 2, most of the stellar
populations in the inner regions of the massive SLACS/LSD early-type
galaxies appear to have been in place already (or was accreted below
that redshift through dry mergers), as indicated by their red colors
and slow evolution at $z \la 1$ (Paper II and KT04).

\noi (2) The dominant old stellar population implies that gas
accretion and subsequent star-formation can only have played a minor
role at lower redshifts (e.g.\ Trager et al.\ 2000; Treu et al.\
2002).  The effect on the inner density slope from adiabatic
contraction, resulting from infall of gas-rich satellites, should
therefore be relatively small as well. If the latter were important
and continued below $z \sim 1$, it could severely affect the evolution
of the logarithmic density slope, which seems to change at most
marginally~(\S\,4).

\noi (3) Once gas was no longer being accreted, the invariance of the
inner density profile (e.g.\ Dehnen 2005; \S\,6.1.1) implies that at
redshifts above 1.5 or 2, the inner regions of these galaxies were
already in dynamical equilibrium (e.g.\ Wechsler et al.\ 2002; Zhao et
al.\ 2003; Gao et al.\ 2004; Kazantzidis et al.\ 2005). Major or minor
collisionless mergers might, however, still occur and replace already
present collisionless matter (Gao et al.\ 2004).

\noi (4) Combining this with the result presented in this paper that
the inner cores are very close to isothermal with little intrinsic
scatter (\S\,3.2; also found by Gerhard et al.\ 2001 at $z \approx 0$)
implies that these cores must have been isothermal already at the time
significant gas-accretion ceased to occur at redshifts of 1.5 or 2 or higher.

\noi (5) The results from numerical simulations (e.g.\ Navarro et al.\
1996; Moore et al.\ 1998), however, suggest that collisionless mergers
in a $\Lambda$CDM cosmology lead to a much more shallow density
profile, never to anything that even remotely resembles an isothermal
profile. We are now faced with a conundrum when trying to explain an
isothermal density profile in a dissipationless scenario, whether the
merging matter is dark or stellar.

\smallskip
This string of arguments suggests that the isothermal nature of the
inner regions of massive early-type galaxies must, somehow, be the
result of the effects of gas-accretion (e.g.\ through mergers) and
subsequent (adiabatic) contraction and star formation. During these
collisional stages in the galaxy-formation process, the sum of the
stellar and dark matter distribution converged to an isothermal
density profile through an as yet unknown process.  Some of the
collisionless matter in the inner regions could be expelled by newly
in-falling dark or stellar matter, redistributing it such that the
phase-space density remains nearly invariant (e.g.\ Gao et al.\ 2004).

This process either occurred rapidly, and only once, for the
early-type galaxy in the ``monolithic'' collapse scenario (e.g.\
Eggen, Lynden-Bell \& Sandage 1962) or in the merging of gas-rich disk
galaxies, {\sl or} it occurred for each of its progenitors, which then
hierarchically and collisionlessly merged. Once the gas supply was
depleted, subsequent dry mergers would retain the isothermality of the
density profile.

The suggested scenario is therefore one where monolithic collapse or
gas-rich disk-galaxy merging occurred for the {\sl progenitors} of
present-day early-type galaxies (only once in the particular case of
the traditional monolithic collapse!) leading to an isothermal density
profile. The resulting gas-poor galaxies subsequently merged
collisionlessly, leaving the density profiles of the merger products
unchanged. Understanding whether this can work, requires detailed
numerical calculations, including baryons and feedback in a realistic
way. These are only recently becoming available (e.g.\ Meza et al.\
2003; Kobayashi 2005) and still include approximations for many
physical processes.

As for observational tests, this hypothetical scenario predicts that
the inner regions of massive early-type galaxies already became
isothermal (in density) at the formation redshift of the old stellar
population. This will be testable if early-type lens galaxies (or
their progenitors) are discovered at redshifts of $z\ga 2$ in future
large-scale surveys e.g.\ with the SKA or the LSST (e.g.\ Koopmans,
Browne \& Jackson 2004) and their density profiles can be quantified
either through lensing or lensing and dynamics combined.

\subsection{Open Questions}

We end with three open and compelling questions as raised by the
results from the SLACS and LSD surveys:

\begin{enumerate}

\item {\sl What is the physical process that leads to an average
isothermal stellar plus dark-matter density profile in the inner
regions of massive early-type galaxies at $z\ga 1$?} The scenario
discussed in \S\,6.2 argues that it most likely is a collisional process.

\item {\sl Why does this process lead to such a remarkably small
scatter in the logarithmic (stellar plus dark-matter) density slope in
their inner regions?}  If collisional processes play a dominant role
in this, as suggested above, it requires strong feedback and an
attractor-like behaviour. Also, subsequent dry mergers can not
increase the scatter significantly (e.g.\ Dehnen 2005), suggesting
that they have very similar isothermal density profiles.

\item {\sl Why does the mass structure in the inner regions of massive
elliptical galaxies evolve so little below a redshift of one?} This
suggests that these galaxies are already dynamically in place at $z
\ga 1$ and that the evolution of their mass structure, through
subsequent merging, plays only a minor role in their inner regions and
must predominantly be dry.

\end{enumerate}

With forthcoming new HST, VLT and Keck data, and improved lensing and
dynamical analysis methods, we soon expect to make further progress in
answering these questions.

{\acknowledgments Based on observations made with the NASA/ESA Hubble
Space Telescope, obtained at the Space Telescope Science Institute,
which is operated by the Association of Universities for Research in
Astronomy, Inc., under NASA contract NAS 5-26555. Support for program
SNAP-10174 was provided by NASA through a grant from the Space
Telescope Science Institute, which is operated by the Association of
Universities for Research in Astronomy, Inc., under NASA contract NAS
5-26555. The authors are grateful for the scheduling work done by
Galina Soutchkova, the Program Coordinator for this HST program.  TT
acknowledges support from NASA through Hubble Fellowship grant
HF-01167.1 and STScI-AR-09222, and thanks UCLA for being such a
welcoming and stimulating Hubble Fellowship host institution during
the initial phases of this project. The authors acknowledge support
from NASA through STScI-AR-09960. The work of LAM was carried out at
Jet Propulsion Laboratory, California Institute of Technology, under a
contract with NASA. LVEK thanks Oleg Gnedin for useful discussions on
adiabatic contraction. The authors thank the referee for helpful
suggestions that further improved the paper.}

\clearpage

\begin{landscape}

\begin{deluxetable}{ccccccccrcccccc}   
  \tablewidth{\vsize}
  \tablecaption{Lensing and Dynamical Model Results}{}  
  \tabletypesize{\scriptsize}
  \startdata
 \hline
 \hline
    Name & $z_{\rm l}$ & $z_{\rm s}$ & ${\rm R}_{\rm e}$ & $\theta_*$  & $q_*$ & $\sigma_{\rm ap}$ & $b_{\rm SIE}$ &
    $q_{\rm SIE}$ & $\theta_{\rm SIE}$ & $\sigma_{\rm SIE}$ & ${\rm R}_{\rm Einst}$  & 
        ${\rm M}_{\rm Einst}$ & $f_{*}$ & $\gamma'$\\
    & & & ($''$) & (deg) & & (km\,s$^{-1}$) & ($''$) & & (deg) & (km\,s$^{-1}$)& (kpc)  & ($10^{10}$~M$_\odot$) \\
    \hline                                                                        
 SDSS J0037$-$0942 & 0.1955 & 0.6322 & 2.38  & 189.5 & 0.76 & 265$\pm$10 &  1.47 & 0.79 & 176.2  &  280 &  4.77 &  27.3  & 0.65$\pm$0.19 &   2.05$\pm$0.07 \\ 
 SDSS J0216$-$0813 & 0.3317 & 0.5235 & 3.37  &  79.2 & 0.85 & 332$\pm$23 &  1.15 & 0.80 &  85.0  &  346 &  5.49 &  48.2  & 0.56$\pm$0.16 &   2.05$\pm$0.21 \\ 
 SDSS J0737+3216   & 0.3223 & 0.5812 & 3.26  & 105.1 & 0.86 & 310$\pm$15 &  1.03 &  0.69 & 100.5  &  297 &  4.83 &  31.2 & 0.63$\pm$0.20 &   2.34$\pm$0.14 \\ 
 SDSS J0912+0029   & 0.1642 & 0.3240 & 4.81  &  13.2 & 0.67 & 313$\pm$12 &  1.61 &  0.56 &   8.7  &  344 &  4.55 &  39.6 & 0.44$\pm$0.13 &   1.82$\pm$0.10 \\ 
 SDSS J0956+5100   & 0.2405 & 0.4700 & 2.60  & 142.0 & 0.76 & 299$\pm$16 &  1.32 &  0.60 & 143.4  &  317 &  5.02 &  37.0 & 0.72$\pm$0.21 &   2.04$\pm$0.12 \\ 
 SDSS J0959+0410  & 0.1260 & 0.5349 & 1.82  &  57.4 & 0.68 & 212$\pm$12 &  1.00 &  0.91 &  71.6  &  216 &  2.25 &   7.7 & 0.79$\pm$0.23 &   2.18$\pm$0.13 \\ 
 SDSS J1250+0523   & 0.2318 & 0.7950 & 1.77  & 110.3 & 0.98 & 254$\pm$14 &  1.15 &  0.97 &  88.7  &  246 &  4.26 &  18.9 & 1.04$\pm$0.30 &   2.26$\pm$0.10 \\ 
 SDSS J1330$-$0148 & 0.0808 & 0.7115 & 1.23  & 103.8 & 0.44 & 178$\pm$9  &  0.85 &  0.70 & 100.0  &  185 &  1.30 &   3.2 & 1.05$\pm$0.30 &   2.18$\pm$0.10 \\
 SDSS J1402+6321   & 0.2046 & 0.4814 & 3.14  &  72.1 & 0.77 & 275$\pm$15 &  1.39 &  0.85 &  62.2  &  298 &  4.66 &  30.3 & 0.82$\pm$0.23 &   1.95$\pm$0.13 \\ 
 SDSS J1420+6019   & 0.0629 & 0.5352 & 2.60  & 110.8 & 0.55 & 194$\pm$5  &  1.04 &  0.73 & 111.7  &  204 &  1.27 &   3.9 & 1.08$\pm$0.31 &   2.03$\pm$0.07 \\ 
 SDSS J1627$-$0053 & 0.2076 & 0.5241 & 2.14  &   5.6 & 0.85 & 275$\pm$12 &  1.21 &  0.92 &  18.7  &  271 &  4.11 &  22.2 & 1.04$\pm$0.30 &   2.21$\pm$0.09 \\ 
 SDSS J1630+4520   & 0.2479 & 0.7933 & 2.02  &  71.7 & 0.83 & 260$\pm$16 &  1.81 &  0.86 &  80.8  &  314 &  7.03 &  50.8 & 0.45$\pm$0.13 &   1.85$\pm$0.10 \\ 
 SDSS J2300+0022   & 0.2285 & 0.4635 & 1.80  &  88.6 & 0.80 & 283$\pm$18 &  1.25 &  0.85 &  94.3  &  302 &  4.56 &  30.4 & 0.75$\pm$0.22 &   2.07$\pm$0.12 \\ 
 SDSS J2303+1422   & 0.1553 & 0.5170 & 4.20  &  38.0 & 0.65 & 260$\pm$15 &  1.64 &  0.62 &  32.5  &  291 &  4.41 &  27.5 & 0.60$\pm$0.17 &   1.82$\pm$0.13 \\ 
 SDSS J2321$-$0939 & 0.0819 & 0.5324 & 4.47  & 126.5 & 0.77 & 236$\pm$7  &  1.57 &  0.82 & 136.2  &  257 &  2.43 &  11.7 & 0.56$\pm$0.16 &   1.87$\pm$0.07 
 \enddata

\tablecomments{\label{tab:results} All position angles are defined
North to East.  The marginalized maximum-likelihood stellar mass
fraction ($f_*$) does not include the prior $f_*\le 1$, which ofcourse
should be satified.  We indicated the maximum-likelihood value, as a
sanity check to show that none of the systems significantly violates
this inequality. Hence the posterior likelihood value, including this
prior, is equal to one, if $f_* > 1$.}
\end{deluxetable}

\clearpage

\end{landscape}


\begin{thebibliography}{}

\bibitem[Abadi et al.(2003)]{2003ApJ...597...21A} Abadi, M.~G., Navarro, 
J.~F., Steinmetz, M., \& Eke, V.~R.\ 2003, \apj, 597, 21 
 
\bibitem[Abadi et al.(2003)]{2003ApJ...591..499A} Abadi, M.~G., Navarro, 
J.~F., Steinmetz, M., \& Eke, V.~R.\ 2003, \apj, 591, 499 
 
\bibitem[Arad \& Lynden-Bell(2005)]{2005MNRAS.361..385A} Arad, I., \& 
Lynden-Bell, D.\ 2005, \mnras, 361, 385 

\bibitem[Arad \& Johansson(2005)]{2005MNRAS.362..252A} Arad, I., \& 
Johansson, P.~H.\ 2005, \mnras, 362, 252 
 
\bibitem[Arnaboldi et al.(1996)]{1996ApJ...472..145A} Arnaboldi, M., et 
al.\ 1996, \apj, 472, 145 

\bibitem[Barnes(1988)]{1988ApJ...331..699B} Barnes, J.~E.\ 1988, \apj, 331, 
699 

\bibitem[Barnes \& Hernquist(1991)]{1991ApJ...370L..65B} Barnes, J.~E., \& 
Hernquist, L.~E.\ 1991, \apjl, 370, L65 

\bibitem[Barnes(1992)]{1992ApJ...393..484B} Barnes, J.~E.\ 1992, \apj, 393, 
484 
 
\bibitem[Barnes \& Hernquist(1992)]{1992ARA&A..30..705B} Barnes, J.~E., \& 
Hernquist, L.\ 1992, \araa, 30, 705 

\bibitem[Barnes \& Hernquist(1996)]{1996ApJ...471..115B} Barnes, J.~E., \& 
Hernquist, L.\ 1996, \apj, 471, 115 

\bibitem[Bell et al.(2004)]{2004ApJ...608..752B} Bell, E.~F., et al.\ 2004, 
\apj, 608, 752 

\bibitem[Bernardi et al.(2003)]{2003AJ....125.1817B} Bernardi, M., et al.\ 
2003, \aj, 125, 1817 

\bibitem[Bertin et al.(1994)]{1994A&A...292..381B} Bertin, G., et al.\ 
1994, \aap, 292, 381 

\bibitem[Binney(1978)]{1978MNRAS.183..501B} Binney, J.\ 1978, \mnras, 183, 
501 

\bibitem[Binney \& Tremaine(1987)]{1987gady.book.....B} Binney, J., \& 
Tremaine, S.\ 1987, Princeton, NJ, Princeton University Press, 1987, 747 
 
\bibitem[Blumenthal et al.(1984)]{1984Natur.311..517B} Blumenthal, G.~R., 
Faber, S.~M., Primack, J.~R., \& Rees, M.~J.\ 1984, \nat, 311, 517 

\bibitem[Blumenthal et al.(1986)]{1986ApJ...301...27B} Blumenthal, G.~R., 
Faber, S.~M., Flores, R., \& Primack, J.~R.\ 1986, \apj, 301, 27 

\bibitem[Bolton \& Burles(2003)]{2003ApJ...592...17B} Bolton, A.~S., \& 
Burles, S.\ 2003, \apj, 592, 17 
 
\bibitem[Bolton et al.(2004)]{2004AJ....127.1860B} Bolton, A.~S., Burles, 
S., Schlegel, D.~J., Eisenstein, D.~J., \& Brinkmann, J.\ 2004, \aj, 127, 
1860 

\bibitem[Bolton et al.(2005)]{2005ApJ...624L..21B} Bolton, A.~S., Burles, 
S., Koopmans, L.~V.~E., Treu, T., \& Moustakas, L.~A.\ 2005, \apjl, 624, 
L21 

\bibitem[Bolton et al.(2006)]{2006ApJ..Bolton} Bolton, A.~S., Burles,
S., Koopmans, L.~V.~E., Treu, T., \& Moustakas, L.~A.\ 2006, \apj,
in press [Paper I]
  
\bibitem[Borriello et al.(2003)]{2003MNRAS.341.1109B} Borriello, A., 
Salucci, P., \& Danese, L.\ 2003, \mnras, 341, 1109 

\bibitem[Boylan-Kolchin et al.(2005)]{2005MNRAS.tmp..655B} Boylan-Kolchin, 
M., Ma, C.-P., \& Quataert, E.\ 2005, \mnras, 655 
 
\bibitem[Brinchmann \& Ellis(2000)]{2000ApJ...536L..77B} Brinchmann, J., \& 
Ellis, R.~S.\ 2000, \apjl, 536, L77 

\bibitem[Bullock et al.(2001)]{2001MNRAS.321..559B} Bullock, J.~S., Kolatt, 
T.~S., Sigad, Y., Somerville, R.~S., Kravtsov, A.~V., Klypin, A.~A., 
Primack, J.~R., \& Dekel, A.\ 2001, \mnras, 321, 559 

\bibitem[Carollo et al.(1995)]{1995ApJ...441L..25C} Carollo, C.~M., de 
Zeeuw, P.~T., van der Marel, R.~P., Danziger, I.~J., \& Qian, E.~E.\ 1995, 
\apjl, 441, L25 

\bibitem[Cohn et al.(2001)]{2001ApJ...554.1216C} Cohn, J.~D., Kochanek, 
C.~S., McLeod, B.~A., \& Keeton, C.~R.\ 2001, \apj, 554, 1216 

\bibitem[Conselice et al.(2003)]{2003AJ....126.1183C} Conselice, C.~J., 
Bershady, M.~A., Dickinson, M., \& Papovich, C.\ 2003, \aj, 126, 1183 

\bibitem[Dalal \& Watson(2004)]{2004astro.ph..9483D} Dalal, N., \& Watson, 
C.~R.\ 2004, submitted to ApJ, astro-ph/0409483

\bibitem[Dehnen(2005)]{2005MNRAS.360..892D} Dehnen, W.\ 2005, \mnras, 360, 
892
 
\bibitem[Dubinski(1994)]{1994ApJ...431..617D} Dubinski, J.\ 1994, \apj, 
431, 617 

\bibitem[Eggen et al.(1962)]{1962ApJ...136..748E} Eggen, O.~J., 
Lynden-Bell, D., \& Sandage, A.~R.\ 1962, \apj, 136, 748 

\bibitem[Eisenstein et al.(2004)]{2004AAS...205.6908E} Eisenstein, D.~J., 
Zehavi, I., Nichol, R., Hogg, D.~W., Blanton, M.~R., Seo, H., Zheng, Z., \& 
Tegmark, M.\ 2004, American Astronomical Society Meeting Abstracts, 205,  

\bibitem[Fabbiano(1989)]{1989ARA&A..27...87F} Fabbiano, G.\ 1989, \araa, 
27, 87 

\bibitem[Fassnacht et al.(1999)]{1999AJ....117..658F} Fassnacht, C.~D., et 
al.\ 1999, \aj, 117, 658 
 
\bibitem[Fassnacht \& Lubin(2002)]{2002AJ....123..627F} Fassnacht, C.~D., 
\& Lubin, L.~M.\ 2002, \aj, 123, 627 

\bibitem[Fischer et al.(1998)]{1998ApJ...503L.127F} Fischer, P., Schade, 
D., \& Barrientos, L.~F.\ 1998, \apjl, 503, L127 
 
\bibitem[Franx et al.(1994)]{1994ApJ...436..642F} Franx, M., van Gorkom, 
J.~H., \& de Zeeuw, T.\ 1994, \apj, 436, 642

\bibitem[Frenk et al.(1985)]{1985Natur.317..595F} Frenk, C.~S., White, 
S.~D.~M., Efstathiou, G., \& Davis, M.\ 1985, \nat, 317, 595

\bibitem[Frenk et al.(1988)]{1988ApJ...327..507F} Frenk, C.~S., White, 
S.~D.~M., Davis, M., \& Efstathiou, G.\ 1988, \apj, 327, 507 
 
\bibitem[Gao et al.(2004)]{2004ApJ...614...17G} Gao, L., Loeb, A., Peebles, 
P.~J.~E., White, S.~D.~M., \& Jenkins, A.\ 2004, \apj, 614, 17 

\bibitem[Gebhardt et al.(2003)]{2003ApJ...597..239G} Gebhardt, K., et al.\ 
2003, \apj, 597, 239 

\bibitem[Gerhard(1981)]{1981MNRAS.197..179G} Gerhard, O.~E.\ 1981, \mnras, 
197, 179 
 
\bibitem[Gerhard et al.(2001)]{2001AJ....121.1936G} Gerhard, O., 
Kronawitter, A., Saglia, R.~P., \& Bender, R.\ 2001, \aj, 121, 1936 

\bibitem[Gnedin et al.(2004)]{2004ApJ...616...16G} Gnedin, O.~Y., Kravtsov, 
A.~V., Klypin, A.~A., \& Nagai, D.\ 2004, \apj, 616, 16 
 
\bibitem[Hernquist(1990)]{1990ApJ...356..359H} Hernquist, L.\ 1990, \apj, 
356, 359 

\bibitem[Hernquist(1992)]{1992ApJ...400..460H} Hernquist, L.\ 1992, \apj, 
400, 460 

\bibitem[Holz(2001)]{2001ApJ...556L..71H} Holz, D.~E.\ 2001, \apjl, 556, 
L71 

\bibitem[Im et al.(2002)]{2002ApJ...571..136I} Im, M., et al.\ 2002, \apj, 
571, 136 
 
\bibitem[Jaffe(1983)]{1983MNRAS.202..995J} Jaffe, W.\ 1983, \mnras, 202, 
995 

\bibitem[Jesseit et al.(2002)]{2002ApJ...571L..89J} Jesseit, R., Naab, T., 
\& Burkert, A.\ 2002, \apjl, 571, L89 

\bibitem[Juneau et al.(2005)]{2005ApJ...619L.135J} Juneau, S., et al.\ 
2005, \apjl, 619, L135

\bibitem[Kauffmann et al.(1996)]{1996MNRAS.283L.117K} Kauffmann, G., 
Charlot, S., \& White, S.~D.~M.\ 1996, \mnras, 283, L117 

\bibitem[Kauffmann \& Charlot(1998)]{1998MNRAS.297L..23K} Kauffmann, G., \& 
Charlot, S.\ 1998, \mnras, 297, L23 

\bibitem[Kawata \& Gibson(2003)]{2003MNRAS.346..135K} Kawata, D., \& 
Gibson, B.~K.\ 2003, \mnras, 346, 135 

\bibitem[Kazantzidis et al.(2004)]{2004ApJ...611L..73K} Kazantzidis, S., 
Kravtsov, A.~V., Zentner, A.~R., Allgood, B., Nagai, D., \& Moore, B.\ 
2004, \apjl, 611, L73 

\bibitem[Keeton et al.(1997)]{1997ApJ...482..604K} Keeton, C.~R., Kochanek, 
C.~S., \& Seljak, U.\ 1997, \apj, 482, 604 
 
\bibitem[Keeton \& Zabludoff(2004)]{2004ApJ...612..660K} Keeton, C.~R., \& 
Zabludoff, A.~I.\ 2004, \apj, 612, 660

\bibitem[Khochfar \& Burkert(2003)]{2003ApJ...597L.117K} Khochfar, S., \& 
Burkert, A.\ 2003, \apjl, 597, L117 

\bibitem[Kochanek(1991)]{1991ApJ...373..354K} Kochanek, C.~S.\ 1991, \apj, 
373, 354 

\bibitem[Kochanek(1994)]{1994ApJ...436...56K} Kochanek, C.~S.\ 1994, \apj, 
436, 56 

\bibitem[Kochanek(1995)]{1995ApJ...445..559K} Kochanek, C.~S.\ 1995, \apj, 
445, 559 
 
\bibitem[Kochanek et al.(2000)]{2000ApJ...543..131K} Kochanek, C.~S., et 
al.\ 2000, \apj, 543, 131 

\bibitem[Kochanek(2002)]{2002ApJ...578...25K} Kochanek, C.~S.\ 2002, \apj, 
578, 25 

\bibitem[Koopmans \& Treu(2002)]{2002ApJ...568L...5K} Koopmans, L.~V.~E., 
\& Treu, T.\ 2002, \apjl, 568, L5 
 
\bibitem[Koopmans \& Treu(2003)]{2003ApJ...583..606K} Koopmans, L.~V.~E., 
\& Treu, T.\ 2003, \apj, 583, 606 
 
\bibitem[Koopmans et al.(2003)]{2003ApJ...595..712K} Koopmans, L.~V.~E., et 
al.\ 2003, \apj, 595, 712 

\bibitem[Koopmans et al.(2003)]{2003ApJ...599...70K} Koopmans, L.~V.~E., 
Treu, T., Fassnacht, C.~D., Blandford, R.~D., \& Surpi, G.\ 2003, \apj, 
599, 70 

\bibitem[Koopmans et al.(2004)]{2004NewAR..48.1085K} Koopmans, L.~V.~E., 
Browne, I.~W.~A., \& Jackson, N.~J.\ 2004, New Astronomy Review, 48, 1085 

\bibitem[Koopmans(2004)]{2004bdmh.confE..66K} Koopmans, L.~V.~E.\
2004, (Electronic) Proceedings of Science, published by SISSA;
Conference: "Baryons in Dark Matter Haloes", Novigrad, Croatia, 5-9
October 2004; editors: R.-J. Dettmar, U. Klein, P. Salucci,
PoS(BDMH2004)066

\bibitem[Koopmans(2005)]{2005MNRAS.363.1136K} Koopmans, L.~V.~E.\ 2005, 
\mnras, 363, 1136 
 
\bibitem[Kormann et al.(1994)]{1994A&A...284..285K} Kormann, R., Schneider, 
P., \& Bartelmann, M.\ 1994, \aap, 284, 285 

\bibitem[Loeb \& Peebles(2003)]{2003ApJ...589...29L} Loeb, A., \& Peebles, 
P.~J.~E.\ 2003, \apj, 589, 29 

\bibitem[Loewenstein \& White(1999)]{1999ApJ...518...50L} Loewenstein, M., 
\& White, R.~E.\ 1999, \apj, 518, 50 

\bibitem[Lynden-Bell(1967)]{1967MNRAS.136..101L} Lynden-Bell, D.\ 1967, 
\mnras, 136, 101 

\bibitem[Ma(2003)]{2003ApJ...584L...1M} Ma, C.\ 2003, \apjl, 584, L1 

\bibitem[Matsushita et al.(1998)]{1998ApJ...499L..13M} Matsushita, K., 
Makishima, K., Ikebe, Y., Rokutanda, E., Yamasaki, N., \& Ohashi, T.\ 1998, 
\apjl, 499, L13 

\bibitem[McIntosh et al.(2005)]{2005ApJ...632..191M} McIntosh, D.~H., et 
al.\ 2005, \apj, 632, 191 

\bibitem[Menanteau et al.(2001)]{2001ApJ...562L..23M} Menanteau, F., 
Jimenez, R., \& Matteucci, F.\ 2001a, \apjl, 562, L23 
 
\bibitem[Menanteau et al.(2001)]{2001MNRAS.322....1M} Menanteau, F., 
Abraham, R.~G., \& Ellis, R.~S.\ 2001b, \mnras, 322, 1 

\bibitem[Meza et al.(2003)]{2003ApJ...590..619M} Meza, A., Navarro, J.~F., 
Steinmetz, M., \& Eke, V.~R.\ 2003, \apj, 590, 619

\bibitem[Mihos \& Hernquist(1996)]{1996ApJ...464..641M} Mihos, J.~C., \& 
Hernquist, L.\ 1996, \apj, 464, 641 

\bibitem[Mo et al.(1998)]{1998MNRAS.295..319M} Mo, H.~J., Mao, S., \& 
White, S.~D.~M.\ 1998, \mnras, 295, 319

\bibitem[Moore et al.(1998)]{1998ApJ...499L...5M} Moore, B., Governato, F., 
Quinn, T., Stadel, J., \& Lake, G.\ 1998, \apjl, 499, L5 

\bibitem[Mould et al.(1990)]{1990AJ.....99.1823M} Mould, J.~R., Oke, J.~B., 
de Zeeuw, P.~T., \& Nemec, J.~M.\ 1990, \aj, 99, 1823 

\bibitem[Mu{\~ n}oz et al.(2001)]{2001ApJ...558..657M} Mu{\~ n}oz, J.~A., 
Kochanek, C.~S., \& Keeton, C.~R.\ 2001, \apj, 558, 657 
 
\bibitem[Naab et al.(2006)]{2006ApJ...636L..81N} Naab, T., Khochfar, S., \& 
Burkert, A.\ 2006, \apjl, 636, L81 

\bibitem[Napolitano et al.(2005)]{2005MNRAS.357..691N} Napolitano, N.~R., 
et al.\ 2005, \mnras, 357, 691 

\bibitem[Navarro \& Benz(1991)]{1991ApJ...380..320N} Navarro, J.~F., \& 
Benz, W.\ 1991, \apj, 380, 320 
 
\bibitem[Navarro et al.(1996)]{1996ApJ...462..563N} Navarro, J.~F., Frenk, 
C.~S., \& White, S.~D.~M.\ 1996, \apj, 462, 563 

\bibitem[Negroponte \& White(1983)]{1983MNRAS.205.1009N} Negroponte, J., \& 
White, S.~D.~M.\ 1983, \mnras, 205, 1009 
 
\bibitem[Nipoti et al.(2003)]{2003MNRAS.342..501N} Nipoti, C., Londrillo, 
P., \& Ciotti, L.\ 2003, \mnras, 342, 501 

\bibitem[Perlmutter et al.(1999)]{1999ApJ...517..565P} Perlmutter, S., et 
al.\ 1999, \apj, 517, 565 

\bibitem[Riess et al.(1998)]{1998AJ....116.1009R} Riess, A.~G., et al.\ 
1998, \aj, 116, 1009 
 
\bibitem[Rix et al.(1997)]{1997ApJ...488..702R} Rix, H., de Zeeuw, P.~T., 
Cretton, N., van der Marel, R.~P., \& Carollo, C.~M.\ 1997, \apj, 488, 702

\bibitem[Romanowsky et al.(2003)]{2003Sci...301.1696R} Romanowsky, A.~J., 
Douglas, N.~G., Arnaboldi, M., Kuijken, K., Merrifield, M.~R., Napolitano, 
N.~R., Capaccioli, M., \& Freeman, K.~C.\ 2003, Science, 301, 1696 

\bibitem[Rusin \& Ma(2001)]{2001ApJ...549L..33R} Rusin, D., \& Ma, C.\ 
2001, \apjl, 549, L33 

\bibitem[Rusin et al.(2002)]{2002MNRAS.330..205R} Rusin, D., Norbury, M., 
Biggs, A.~D., Marlow, D.~R., Jackson, N.~J., Browne, I.~W.~A., Wilkinson, 
P.~N., \& Myers, S.~T.\ 2002, \mnras, 330, 205 

\bibitem[Rusin et al.(2003)]{2003ApJ...587..143R} Rusin, D., et al.\ 2003, 
\apj, 587, 143

\bibitem[Rusin et al.(2003)]{2003ApJ...595...29R} Rusin, D., Kochanek, 
C.~S., \& Keeton, C.~R.\ 2003, \apj, 595, 29 

\bibitem[Rusin \& Kochanek(2005)]{2005ApJ...623..666R} Rusin, D., \& 
Kochanek, C.~S.\ 2005, \apj, 623, 666 

\bibitem[Ryden \& Gunn(1987)]{1987ApJ...318...15R} Ryden, B.~S., \& Gunn, 
J.~E.\ 1987, \apj, 318, 15 
  
\bibitem[Saglia et al.(1992)]{1992ApJ...384..433S} Saglia, R.~P., Bertin, 
G., \& Stiavelli, M.\ 1992, \apj, 384, 433 

\bibitem[Schneider et al.(1992)]{1992grle.book.....S} Schneider, P.,
Ehlers, J., \& Falco, E.~E.\ 1992, Gravitational Lenses,
Springer-Verlag Berlin Heidelberg New York.

\bibitem[Schweizer(1982)]{1982ApJ...252..455S} Schweizer, F.\ 1982, \apj, 
252, 455

\bibitem[Seljak(2002)]{2002MNRAS.334..797S} Seljak, U.\ 2002, \mnras, 334, 
797 

\bibitem[Sheth et al.(2003)]{2003ApJ...594..225S} Sheth, R.~K., et al.\ 
2003, \apj, 594, 225 

\bibitem[Spergel et al.(2003)]{2003ApJS..148..175S} Spergel, D.~N., et al.\ 
2003, \apjs, 148, 175 

\bibitem[Stiavelli \& Bertin(1987)]{1987MNRAS.229...61S} Stiavelli, M., \& 
Bertin, G.\ 1987, \mnras, 229, 61 
 
\bibitem[Tegmark et al.(2004)]{2004PhRvD..69j3501T} Tegmark, M., et al.\ 
2004, \prd, 69, 103501 

\bibitem[Toomre \& Toomre(1972)]{1972ApJ...178..623T} Toomre, A., \& 
Toomre, J.\ 1972, \apj, 178, 623 

\bibitem[Trager et al.(2000)]{2000AJ....120..165T} Trager, S.~C., Faber, 
S.~M., Worthey, G., \& Gonz{\' a}lez, J.~J.\ 2000, \aj, 120, 165 

\bibitem[Tran et al.(2005)]{2005ApJ...627L..25T} Tran, K.-V.~H., van 
Dokkum, P., Franx, M., Illingworth, G.~D., Kelson, D.~D., \& Schreiber, 
N.~M.~F.\ 2005, \apjl, 627, L25

\bibitem[Treu et al.(2002)]{2002ApJ...564L..13T} Treu, T., Stiavelli, M., 
Casertano, S., M{\o}ller, P., \& Bertin, G.\ 2002, \apjl, 564, L13 

\bibitem[Treu \& Koopmans(2002)]{2002MNRAS.337L...6T} Treu, T., \& 
Koopmans, L.~V.~E.\ 2002, \mnras, 337, L6 

\bibitem[Treu \& Koopmans(2002)]{2002ApJ...575...87T} Treu, T., \& 
Koopmans, L.~V.~E.\ 2002, \apj, 575, 87 

\bibitem[Treu \& Koopmans(2003)]{2003MNRAS.343L..29T} Treu, T., \& 
Koopmans, L.~V.~E.\ 2003, \mnras, 343, L29 
 
\bibitem[Treu \& Koopmans(2004)]{2004ApJ...611..739T} Treu, T., \& 
Koopmans, L.~V.~E.\ 2004, \apj, 611, 739 

\bibitem[Treu et al.(2005a)]{2005ApJ...622L...5T} Treu, T., Ellis, R.~S., 
Liao, T.~X., \& van Dokkum, P.~G.\ 2005a, \apjl, 622, L5 

\bibitem[Treu et al.(2005b)]{2005ApJ...633} Treu, T., et al. 2005b,
\apj, in press

\bibitem[Treu et al.(2006)]{2006ApJ...633} Treu, T., Koopmans,
L.V.E., Bolton, A.S., Burles, S., Moustakas, L.A. 2006, ApJ,
in press [Paper II]

\bibitem[van Albada(1982)]{1982MNRAS.201..939V} van Albada, T.~S.\ 1982, 
\mnras, 201, 939 
 
\bibitem[van Dokkum et al.(1999)]{1999ApJ...520L..95V} van Dokkum, P.~G., 
Franx, M., Fabricant, D., Kelson, D.~D., \& Illingworth, G.~D.\ 1999, 
\apjl, 520, L95 

\bibitem[van Dokkum \& Franx(2001)]{2001ApJ...553...90V} van Dokkum, P.~G., 
\& Franx, M.\ 2001, \apj, 553, 90 

\bibitem[Warren \& Dye(2003)]{2003ApJ...590..673W} Warren, S.~J., \& Dye, 
S.\ 2003, \apj, 590, 673 

\bibitem[Wechsler et al.(2002)]{2002ApJ...568...52W} Wechsler, R.~H., 
Bullock, J.~S., Primack, J.~R., Kravtsov, A.~V., \& Dekel, A.\ 2002, \apj, 
568, 52 

\bibitem[White \& Frenk(1991)]{1991ApJ...379...52W} White, S.~D.~M., \& 
Frenk, C.~S.\ 1991, \apj, 379, 52 
 
\bibitem[Winn et al.(2003)]{2003ApJ...587...80W} Winn, J.~N., Rusin, D., \& 
Kochanek, C.~S.\ 2003, \apj, 587, 80 

\bibitem[Wucknitz(2002)]{2002MNRAS.332..951W} Wucknitz, O.\ 2002, \mnras, 
332, 951 

\bibitem[Wucknitz et al.(2004)]{2004MNRAS.349...14W} Wucknitz, O., Biggs, 
A.~D., \& Browne, I.~W.~A.\ 2004, \mnras, 349, 14 

\bibitem[Zhao et al.(2003)]{2003MNRAS.339...12Z} Zhao, D.~H., Mo, H.~J., 
Jing, Y.~P., B\"orner, G.\ 2003, \mnras, 339, 12 
 
\end{thebibliography}
\end{document}